\documentclass{ptephy}

\newcommand{\Slash}[1]{{\ooalign{\hfil$#1$\hfil\crcr\raise.167ex\hbox{/}}}}
\newcommand{\ltsim}{\protect\raisebox{-0.5ex}{$\:\stackrel{\textstyle <}{\sim}\:$}}

\usepackage{graphicx}
\usepackage{wrapft}
\usepackage{lscape}

\begin{document}

\title{Hadron-Quark Crossover and Massive Hybrid Stars}

\author{\name{Kota Masuda}{1,2,\ast}, \name{Tetsuo Hatsuda}{2}, and \name{Tatsuyuki Takatsuka}{3}\thanks{These authors contributed equally to this work.}}

\address{\affil{1}{Department of Physics, The University of Tokyo, Tokyo 113-0033, Japan}
\affil{2}{Theoretical Research Division, Nishina Center, RIKEN, Wako 351-0198, Japan}
\affil{3}{Iwate University, Morioka 020-8550, Japan}
\email{masuda@nt.phys.s.u-tokyo.ac.jp}}

\begin{abstract}
 On the basis of the percolation picture  from the hadronic phase with hyperons to the quark phase with strangeness, we construct a new equation of state (EOS) with the pressure  interpolated
 as a function of the baryon density.
  The maximum mass of neutron stars can exceed $2M_{\odot}$
    if the following two conditions are satisfied;  
(i) the crossover from the hadronic matter to the quark matter takes place
at around three times the normal nuclear matter density, and 
 (ii) the quark matter is strongly interacting in the crossover region and 
 has stiff equation of state.
 This is in contrast to the conventional approach assuming the 
  first order phase transition in which the EOS becomes always soft 
   due to the presence of the quark matter at high density.
  Although the choice of the hadronic EOS does not affect the above conclusion on the 
  maximum mass,  the three-body force among nucleons and hyperons
  plays an essential role for the onset of the hyperon mixing and the
  cooling of neutron stars.
\end{abstract}

\subjectindex{Neutron stars, Nuclear matter aspects in nuclear astrophysics, Hadrons and quarks in nuclear matter, Quark matter}

\maketitle

\section{Introduction}

Neutron star (NS)
is a cosmic laboratory which provides us with a testing ground for
the rich phase structure of quantum chromodynamics (QCD) \cite{fukushima-hatsuda}
through the observables such as the mass ($M$), the
radius ($R$), the surface temperature ($T_s$), the surface
magnetic field ($B_s$) and so on \cite{NS-review}.
Among others,
$M$ and $R$ are particularly important probes for constraining
the equation of state (EOS) and the composition of high density matter.

From the theoretical point of view, the onset of the strangeness degrees
of freedom
inside the NSs has attracted much attention in recent years:
General consensus is that the hyperons ($Y$)
such as $\Lambda$ and $\Sigma^-$ would participate in NS cores
at densities of several times nuclear matter density
($\rho_0=0.17$ fm$^{-3}$)\cite{takatsuka_2004,Weissenborn_2010,
shulze-2006,30,Ozel_2010,djapo-2010,Kurkela}.
The precise value of the threshold density $\rho_{\rm th}$ depends on the
hyperon-nucleon interactions which have still uncertainties at the moment
but will be improved by the future hypernuclear data \cite{Nagae,Tamura,Nakazawa}
and by the lattice QCD simulations \cite{HAL}.
From the observational point of view,
a massive NS, PSR J1614-2230, with $M_{\rm{obs}} = (1.97\pm0.04)M_{\odot}$
was recently discovered \cite{PSR}.
 Conflict between the $2M_{\odot}$-NS which requires stiff EOS
and the $Y$-mixing which gives soft EOS leads to a challenging problem whether
massive neutron stars are in contradiction to the existence of the
exotic components such as the hyperons and deconfined quarks\cite{Kim,weissenborn-2011,
Klahn,Weissenborn,Bonanno,Chen,Schramm,Whittenbury,saito}.

The purpose of the present paper is to investigate whether the
``hybrid stars" which have quark matter in the core are compatible with
$2M_{\odot}$-NS.
Historically, the transition from the hadronic matter to the quark matter
has been assumed to be the first-order phase transition and the
Gibbs phase equilibrium conditions  are imposed.
However, treating the point-like hadron as
an independent degree of freedom is not fully justified in the transition region
because all hadrons are extended objects composed of quarks and gluons.
Furthermore, the system must be strongly interacting
in the transition region, so that it  can be described
neither by an extrapolation of the hadronic EOS
from the low-density side nor by an extrapolation of the quark EOS
from the high-density side \cite{takatsuka-2011}. This is analogous to
the BEC-BCS crossover realized in the many-body system of
ultra-cold fermionic atoms \cite{BEC-BCS}.

\begin{figure}[!b]
\begin{center}
\centerline{\includegraphics[scale=0.5]{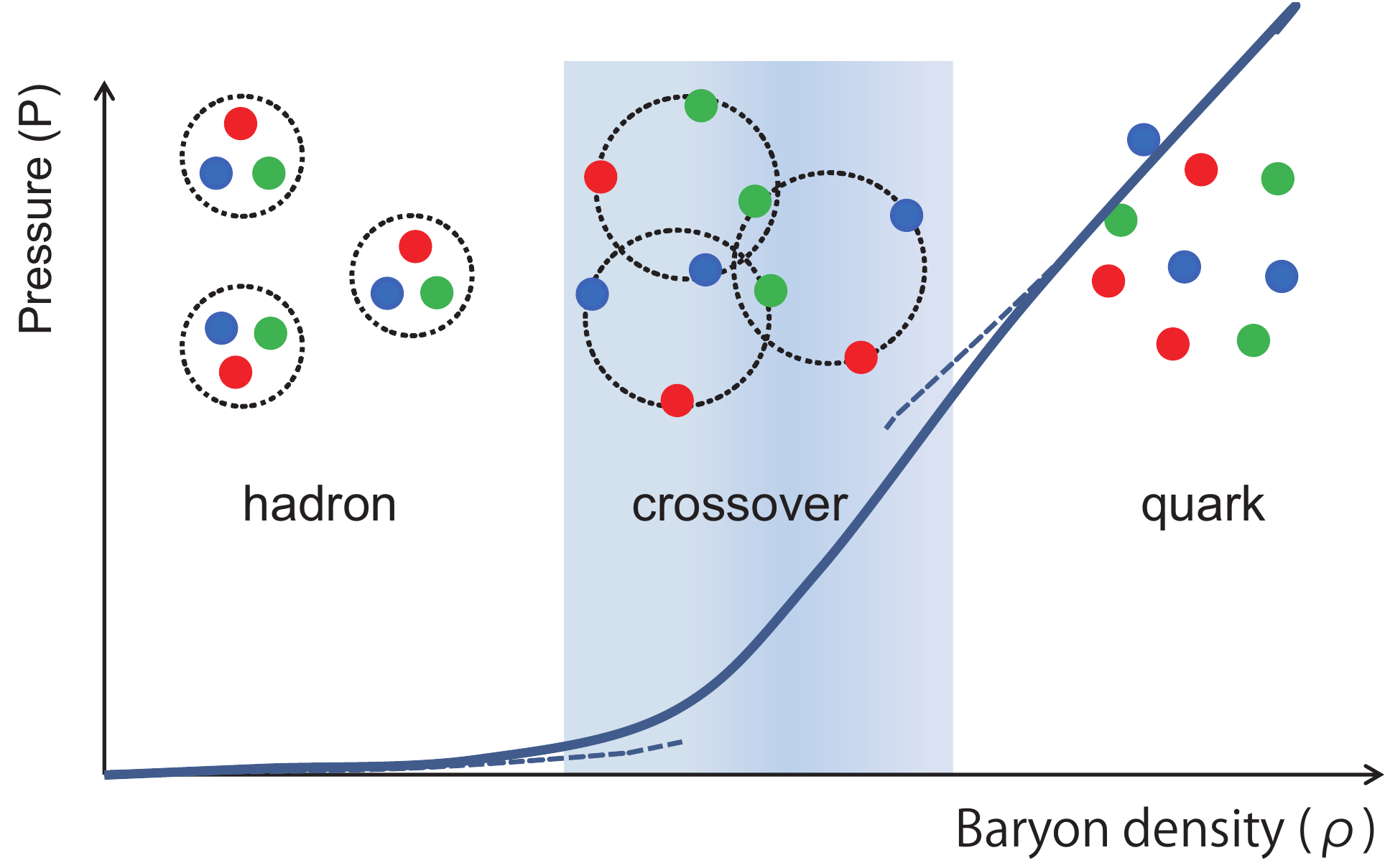}}
\end{center}
\caption{\footnotesize{Schematic picture of the QCD pressure ($P$) as a function of 
 the baron density ($\rho$) under the assumption of the hadron-quark crossover.
 The crossover region where finite-size hadrons start to overlap and percolate
   is shown by the shaded area.  The pressure calculated on the basis 
   of the point-like hadrons (shown by the dashed line at low density)
    and that calculated on the basis of 
    weakly interacting quarks (shown by the dashed line at high density)
    lose their validity in the crossover region, so that
    the naive use of the Gibbs conditions by extrapolating the 
    dashed lines is not  justified in general.  }}
\label{crossover-1}
\end{figure}

Fig.~\ref{crossover-1} illustrates the above situation
in terms of the pressure as a function of baryon density ($\rho$).
One may expect a gradual onset of quark degrees of freedom in dense
matter associated with
the percolation of finite size hadrons, i.e., a smooth crossover from
the hadronic
matter to the quark matter. Such a percolation picture of hadrons
has been discussed in seminal works such as Refs.\cite{baym,satz}.
Also, hadron-quark continuity \cite{Schafer-wilczek,Fukushima}
and hadron-quark crossover \cite{hatsuda-baym,Maeda}
have been discussed in relation to the existence of 
  color superconductivity at high density. 
  In this paper, we show that
 the crossover picture can lead to a stiffening of EOS 
unlike the case of the first-order transition, if the following conditions are met:
  (i) the crossover 
  takes place at relatively low density (around three times the normal
nuclear matter density), and (ii) the strongly interacting quark matter has stiff EOS.
 This implies that the hadron-quark crossover provides us
with a novel mechanism to support  massive neutron stars with 
 quark core. Preliminary account of our results has been reported in \cite{masuda}.
 We note that an interpolation between hadronic matter and stiff quark matter
  was previously considered phenomenologically in \cite{Kalogera}. 
 
 This paper is organized as follows.
In \S 2, the characteristic features of the
hadronic EOSs (H-EOSs) to be used at low densities are summarized.
In \S 3, we treat the strongly interacting quark matter by using the
Nambu-Jona-Lasinio (NJL) type model and derive the quark EOS (Q-EOS) to
be used at high densities.
In \S 4, we describe our interpolation procedure to obtain the EOS 
 in the hadron-quark crossover region.
In \S 5, numerical results and discussions are given for the bulk
properties of hybrid stars,
such as the $M-R$ relationship, the maximum mass $M_{\rm{max}}$ and
the $M-\rho_c ({\rm central\ density})$ relationship.
We discuss how these results depend on the different choice of H-EOS and Q-EOS.
 A comment on the cooling of NSs with respect to the
hyperon mixture inside the core is also given.
\S 6 is devoted to concluding remarks.

\section{Hadronic EOS (H-EOS)}

We consider several different EOSs  with $Y$-mixing:
\begin{itemize}
  \item TNI2, TNI3, TNI2u and TNI3u  \cite{Nishizaki_2001,Nishizaki_2002}: 
  TNI2 and TNI3 are obtained by the G-matrix calculation
  with  Reid soft-core potential for $NN$ and
 Nijmegen type-D hard-core potential for $YN$ and $YY$.
 Also, a phenomenological three-body force   \cite{Friedman}
 is  introduced  in a form of effective $NN$ force to reproduce the saturation point of 
  symmetric nuclear matter with the  incompressibility $\kappa$=250MeV 
  (TNI2) and $\kappa$=300MeV (TNI3).
  For TNI2u and TNI3u, the three-body interaction is introduced {\it universally}
   in a form of effective $NN$, $NY$ and $YY$ forces.
\item AV18+TBF and Paris+TBF \cite{Baldo}:
 They are obtained  by the
G-matrix calculation but with different choice of potentials, 
AV18 and  Paris  potentials for $NN$ and
Nijmegen soft-core potential for $YN $ and $YY$.
Also, the three-body force of Urbana-type is introduced in a form of effective $NN$ force 
to meet the saturation condition. 
\item  SCL3$\Lambda\Sigma$ \cite{Tsubakihara}: It is based on a relativistic
mean field (RMF) model with  chiral SU(3) symmetry and 
logarithmic potential motivated by the strong coupling lattice QCD approach.
 Phenomenological parameters of the model are determined 
 to reproduce the saturation condition, bulk properties of normal nuclei
and separation energies of single- and double-$\Lambda$ hypernuclei.
\end{itemize}

\begin{figure}[!b]
  \begin{center}
      \centerline{\includegraphics[scale=0.55]
                                     {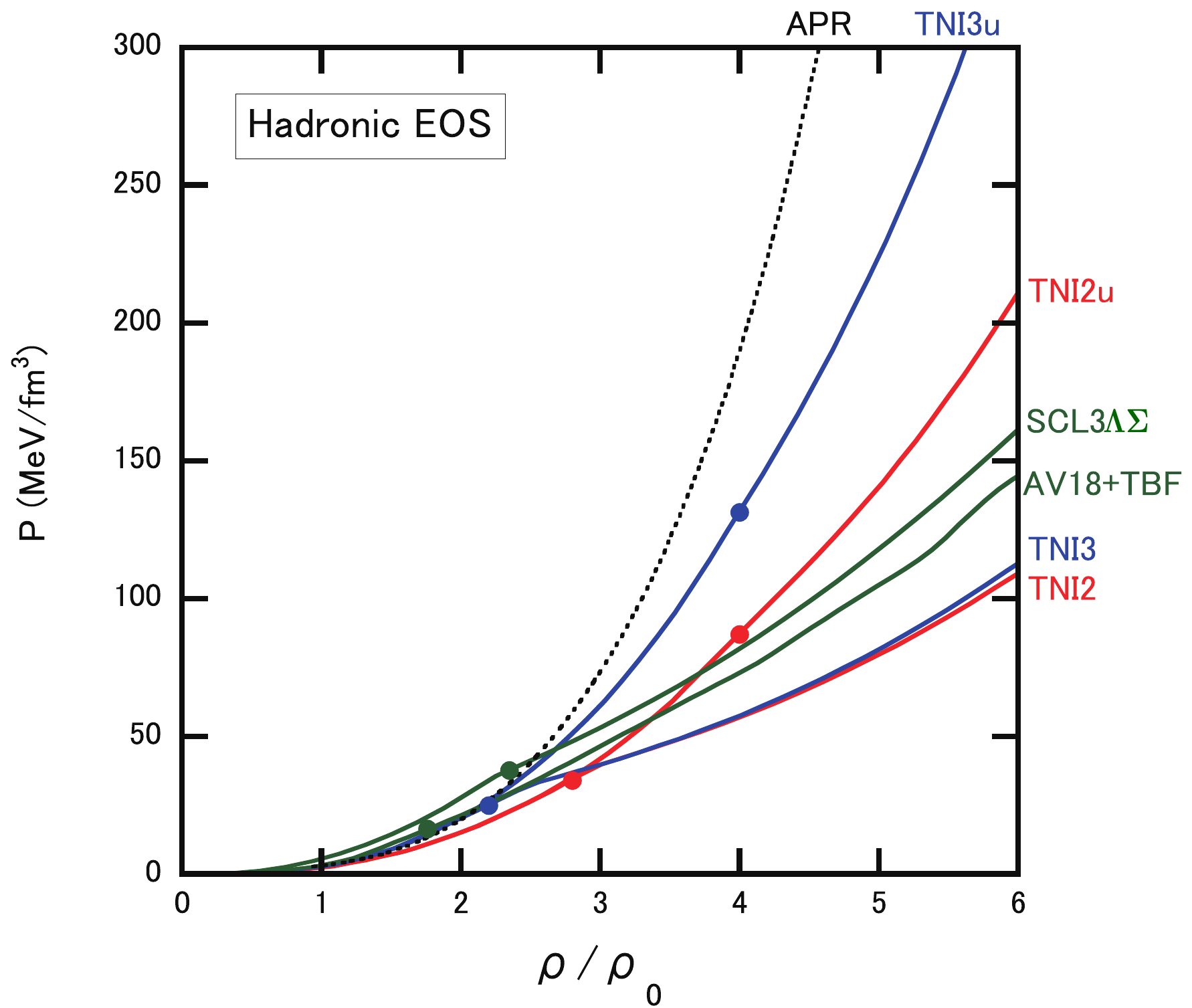}}
\end{center}
      \caption{\footnotesize{ 
Pressure $(P)$ for the $Y$-mixed neutron star matter with $\beta$-equilibrium and charge neutrality as a function of the total baryon density $\rho$ for different types of EOS. 
 Solid red lines: TNI2u (G-matrix approach, universal three-body force, $\kappa=250$MeV) 
 and TNI2 (G-matrix approach, three-nucleon force, $\kappa=250$MeV).  
 Solid blue lines: TNI3u (G-matrix approach, universal three-body force, $\kappa=300$MeV) 
 and TNI3 (G-matrix approach, three-nucleon force, $\kappa=300$MeV) \cite{Nishizaki_2001,Nishizaki_2002}.
 Solid green lines: AV18+TBF (G-matrix approach, three-nucleon force, $\kappa=192$MeV) \cite{Baldo} 
 and SCL3$\Lambda \Sigma$ (relativistic mean field model with chiral SU(3) symmetry, $\kappa=211$MeV)
  \cite{Tsubakihara}. 
 Paris+TBF is not plotted here because it is almost the same as AV18+TBF. 
  For comparison, $P$ for the neutron star matter without hyperons obtained from
  APR EOS \cite{Akmal:1998}  is also plotted  by the dotted lines. 
}}
      \label{PwithY}
\end{figure}

\begin{figure}[!t]
  \begin{center}
      \centerline{\includegraphics[scale=0.55]
                                     {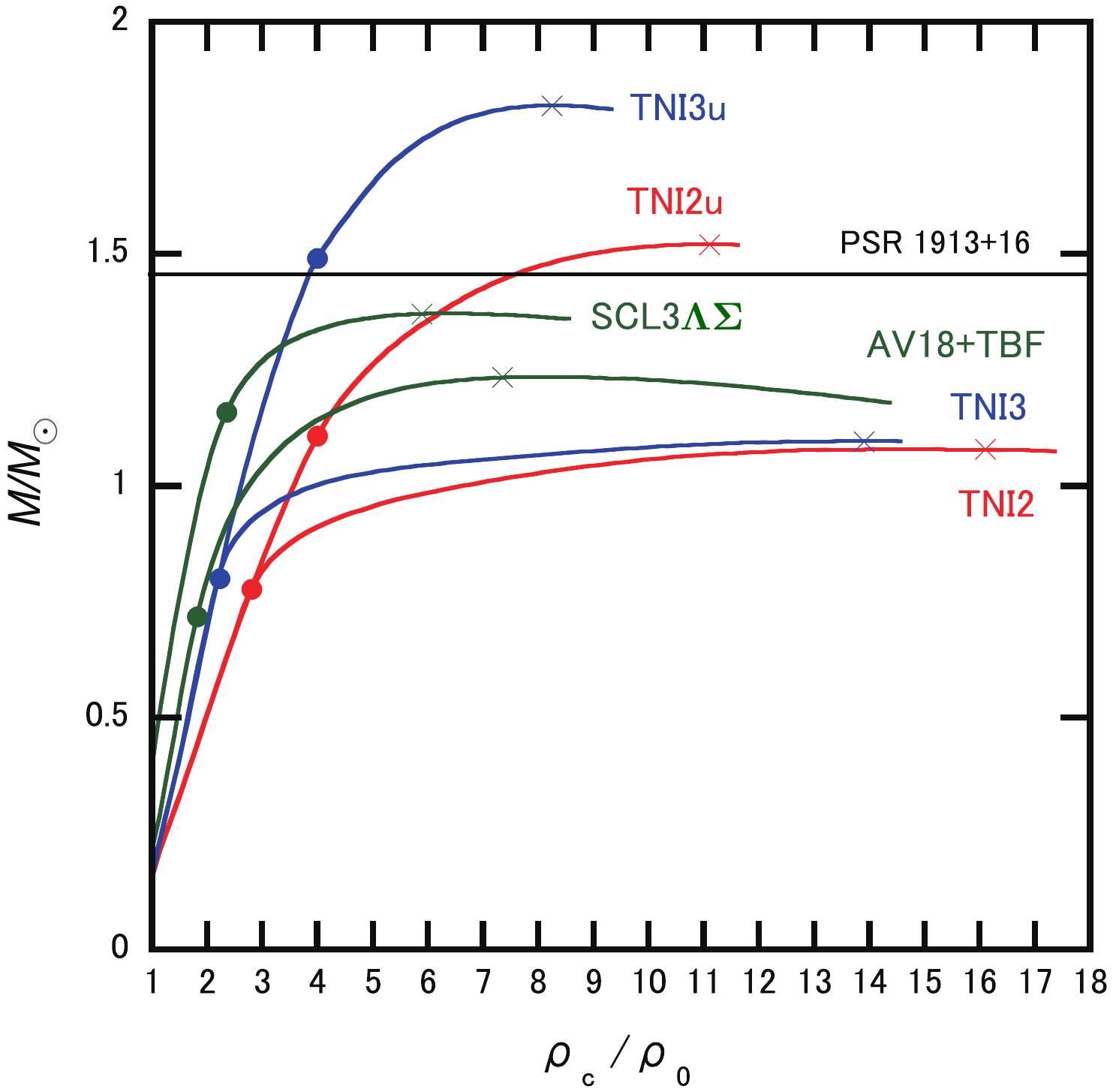}}
\end{center}
      \caption{\footnotesize{$M$-$\rho_c$ relationship
      corresponding to EOSs in Fig.\ref{PwithY} (details of EOS
are given in Table A1 of Appendix A). Colors on each line are the same with those in Fig.\ref{PwithY}.      
      The cross symbols denote the points where the NS mass becomes maximum, $M_{\rm{max}}$.
Solid black line denotes $M=1.44M_{\odot}$ for PSR 1913+16.}}
      \label{M-rho-hadron}
\end{figure}

In Fig.\ref{PwithY}, we plot the pressure $P$ for the $Y$-mixed 
neutron star matter with $\beta$-equilibrium and 
 charge neutrality  as a function of baryon density $\rho$ obtained
 from the EOSs listed above (Paris+TBF is not shown since it is almost the same as AV18+TBF). 
 For comparison, $P$ for the neutron star matter without hyperons obtained from
  APR EOS (\cite{Akmal:1998})  is also plotted in Fig.\ref{PwithY} by the dotted lines. 
In Fig. \ref{M-rho-hadron}, the $M-\rho_c$ relationships for corresponding NS models are shown.
The filled circle on each curve denotes the threshold density of 
 $Y$-mixture. There are some features to be remarked from
  the figure:  (i)  Different H-EOSs do not show significant 
  difference in $P$ up to 2.5 $\rho_0$, and (ii) 
  the $Y$-mixture is delayed from
   (2-3)$\rho_0$ to 4$\rho_0$ if there exits
  repulsive three-body force universally for baryons as in the case of TNI2u and TNI3u.
Even light-mass NSs ($M<M_{\odot}$ for TNI2, TNI3 and AV18+TBF and $M<1.2M_{\odot}$ for SCL3$\Lambda \Sigma$) have already the $Y$-mixed core.

\begin{table}[!t]
\caption{\footnotesize{Properties of various hadronic EOSs with hyperons;  
  TNI2, TNI3, TNI2u, TNI3u \cite{Nishizaki_2001,Nishizaki_2002}, Paris+TBF, AV18+TBF \cite{Baldo,Li,Schulze} and SCL3$\Lambda \Sigma$ \cite{Tsubakihara}.  $\kappa$ is the nuclear incompressibility and $\rho_{\rm th}$ is the threshold density of 
  hyperon-mixing with $\rho_0$ (=0.17/fm$^3$) being the normal nuclear density. $R$ and $\rho_c$ denote
the radius and central density for the maximum mass ($M_{\rm max}$) NS, respectively.
 The numbers in 
 the parentheses are those without hyperons. $\ast$s indicate that
 the numbers are read from the figures in \cite{Baldo}.} }
\label{property of H-EOS}
\begin{center}
  \begin{tabular}[c]{c|c|c|c|c|c|c|c}  \hline \hline
  EOS & TNI2 & TNI3 & TNI2u & TNI3u & Paris+TBF & AV18+TBF & SCL3$\Lambda \Sigma$ \\ \hline
$\kappa$ (MeV) & 250 & 300  & 250 & 300 & 281  & 192 & 211        \\ 
$\rho_{\rm th}(\Lambda)/\rho_0$  & 2.95 & 2.45  & 4.01 & 4.01 & 2.9$^{\ast}$   & 2.8$^{\ast}$  & 2.24     \\ 
$\rho_{\rm th}(\Sigma^-)/\rho_0$ & 2.83 & 2.23  & 4.06 & 4.01 & 1.9$^{\ast}$   & 1.8$^{\ast}$  & 2.24     \\ \hline
$M_{\rm{max}}/M_{\odot}$  & 1.08 & 1.10  & 1.52 & 1.83 & 1.26  & 1.22  & 1.36     \\
                          &(1.62)&(1.88) &      &      &(2.06) &(2.00) &(1.65)    \\
R(km)                     & 7.70 & 8.28  & 8.43 & 9.55 & 10.46 & 10.46 & 11.42    \\
                          &(8.64)&(9.46) &      &      &(10.50)&(10.54)&(10.79)   \\
$\rho_c/\rho_0$           & 16.10& 13.90 &11.06 & 8.26 & 7.35  &7.35   & 6.09     \\
                          &(9.97)&(8.29) &      &      &(6.47) &(6.53) &(6.85)    \\
\hline \hline
\end{tabular} 
\end{center}
\end{table}

    In Table \ref{property of H-EOS}, we show  $\kappa$
 and the threshold densities of hyperon$-$mixing, $\rho_{\rm th}(\Lambda)$ and 
 $\rho_{\rm th}(\Sigma^-)$, for each H-EOS.
 In the same table, we show the maximum-mass $M_{\rm max}$, the radius
  $R$ and the central density $\rho_{\rm c}$ of the NS obtained from each H-EOS.
   The values obtained by switching off the 
   $Y$-mixing are  given in the parentheses for comparison.
 For the H-EOSs without universal three-body repulsion, significant 
softening due to $Y$-mixing reduces $M_{\rm max}$, i.e.,
 $M_{\rm max}$ (without $Y$)=$(1.62-2.00)$$\rightarrow$
$M_{\rm max}$ (with $Y$)=$(1.08-1.26)$.
 This clearly contradicts the observed mass $M_{\rm obs}=1.44M_{\odot}$
for PSR1913+16. On the other hand, H-EOSs with universal three-body repulsion
 (TNI2u, TNI3u), $M_{\rm max}$ is recovered nearly to that without $Y$.

The use of several kinds of EOS mentioned above, from different
theoretical methods (G-matrix, RMF),
 with various stiffness ranging from $\kappa \sim 190$ MeV to
300 MeV and with the variation of $\rho_{\rm th}(Y) \simeq (2-4) \rho_0$,
 is expected to cover the present uncertainties of the H-EOSs.
For completeness, 
numerical values of the pressure $P$ and the energy density $\varepsilon$ 
as a function of the baryon density
are tabulated  in Table A1 in  Appendix A.

\section{Quark EOS (Q-EOS)}

The baryon density at the central core of the NSs would be 
at most 10$\rho_0$.  Although hadrons do not keep their identities
 in such a high density, the chemical potential of the quarks are
 about $(400-500)$MeV which is not  high enough for the asymptotic freedom
 at work. Namely, the deconfined quarks inside the NSs, even if they exist,
 would be strongly interacting.  
Analogous situation at finite 
temperature 
has been  expected theoretically and is recently confirmed
by  the relativistic heavy-ion collisions at RHIC and LHC;
 it is now called the  strongly interacting quark-gluon plasma (sQGP).

 Since lattice QCD
 to treat the strongly interacting quark matter (sQM) at finite baryon density
  is unfortunately not possible due to the notorious sign problem, 
 we adopt an effective theory of QCD, the (2+1)-flavor Nambu$-$Jona-Lasinio (NJL) model.
 This model is particularly useful to take into account 
 the important phenomena such as the 
  partial restoration  of chiral symmetry at high density 
  \cite{Vogel,Klevansky,Hatsuda-Kunihiro,Buballa}. 
 
  The model Lagrangian we consider is
\begin{eqnarray}
{\mathcal L}_{\rm NJL}
&=&\overline{q}(i \Slash \partial-m)  q+\frac{1}{2}G_{_S}\sum_{a=0}^{8}[(\overline{q}\lambda^a q)^2+(\overline{q}i\gamma_5\lambda^a q)^2]
- G_{_D}[\mathrm{det}\overline{q}(1+\gamma_5) q+ {\rm h.c.}] \nonumber \\
&& -\begin{cases}
\frac{1}{2}g_{_V}(\overline{q}\gamma^{\mu}q)^2 \\
\frac{1}{2}G_{_V}\sum_{a=0}^8\left[(\overline{q}\gamma^{\mu}\lambda^a q)^2
+(\overline{q}i\gamma^{\mu}\gamma_5 \lambda^a q)^2\right]\ 
\end{cases}
\label{eq-1}
\end{eqnarray}
where the quark field $q_i$ ($i=u,d,s$) has three colors and  three flavors with
the current quark mass $m_i$.
The term proportional to $G_{_S}$ is a $U(3)_L \times U(3)_R$ symmetric
four-fermi interaction where  $\lambda^a$ are the Gell-Mann matrices with
$\lambda^0=\sqrt{2/3}\ {\rm I}$. 
The  term proportional to $G_{_D}$ is the 
Kobayashi$-$Maskawa$-$'t Hooft (KMT)
six-fermi interaction which breaks $U(1)_A$ symmetry.
 We consider two types of vector interaction (the second line of 
Eq.(\ref{eq-1})):
The term proportional to $g_{_V} (>0)$ gives a universal repulsion among different flavors,
 while the one proportional to $G_{_V} (>0)$ gives  flavor-dependent repulsion.

 In the mean-field approximation, the constituent quark masses $M_i$ ($i=u,d,s$) 
are generated  dynamically through the NJL interactions ($G_{S,D}$), 
\begin{eqnarray}  
M_i=m_i-2G_{_S} \sigma_i +2G_{_D} \sigma_j \sigma_k,
\end{eqnarray}  
where $\sigma_i = \langle \bar{q}_iq_i \rangle$
is the quark condensate in each flavor, and ($i$, $j$, $k$) corresponds to  the cyclic 
permutation of $u, d$ and $s$. The thermodynamic potential $\Omega$ is related to the pressure
 as $\Omega=-T\rm{log}Z = -P V$, so that we have
\begin{eqnarray}
P(T,\mu_{u,d,s})&=&
T\sum_i \sum_{\ell} \int \frac{d^3p}{(2\pi)^3}\mathrm{Trln}
\left(\frac{S_i^{-1}(i\omega_{\ell},{\bf{p}})}{T}\right)
\nonumber \\
&&-G_{_S}\sum_{i}\sigma_{i}^2-4G_{_D}\sigma_{u}\sigma_{d} \sigma_{s} 
+ \begin{cases}
\frac{1}{2}g_{_V}\left( \sum_i n_i \right)^2 \\
\frac{1}{2}G_{_V} \sum_i n_i^2 
\end{cases}
\label{eq:NJL-pressure}
\end{eqnarray}
where $n_i = \langle q^{\dagger}_iq_i \rangle$ is the quark number density in each flavor,
 and  $S_i$ is the quark propagator, which can be written as
\begin{eqnarray}
S_i^{-1}=\Slash p-M_i-\gamma^0 \mu_i^{\rm{eff}},
\ \ \ \mu_i^{\rm{eff}}
\equiv \begin{cases}
\mu_i-g_{_V}\sum_j n_j \\
\mu_i-G_{_V} n_i 
\end{cases}
\label{eq:S-prop}
\end{eqnarray}
where $i\omega_{\ell}=(2\ell +1)\pi T$ and $\mu_i^{\rm{eff}}$ is an effective chemical potential \cite{Asakawa-Yazaki}.

There are six independent 
parameters in the (2+1)-flavor NJL model; the UV cutoff, $\Lambda$, the 
coupling constants, $G_{_S},G_{_D}$ and $g_{_V} (G_{_V})$, and the quark masses,
 $m_{u,d}$ and $m_{s}$.  Five parameters except for $g_{_V} (G_{_V})$ 
 have been determined from hadron phenomenology.  We consider three parameter
 sets summarized in Table \ref{table NJL};
  HK (Hatsuda and Kunihiro), RKH (Rehberg, Klevansky and Hufner)
 and LKW (Lutz, Klimt and Weise)   \cite{Vogel,Klevansky,Hatsuda-Kunihiro,Buballa}. 

\begin{table}[!h]
\caption{\footnotesize{Parameter sets of (2+1)-flavor NJL model \cite{Vogel,Klevansky,Hatsuda-Kunihiro,Buballa}.}} 
\label{table NJL}
\begin{center}
  \begin{tabular}[c]{c|c|c|c|c|c}  \hline \hline
     
 & $\Lambda$(MeV) 
 & $G_{_S}\Lambda^2$ 
 & $G_{_D}\Lambda^5$ 
 & $m_{u,d}$(MeV)
 & $m_s$(MeV)   \\ \hline 
 HK   & 631.4 & 3.67 & 9.29  & 5.5 & 135.7    \\ 
 RKH  & 602.3 & 3.67 & 12.36 & 5.5 & 140.7    \\ 
 LKW  & 750   & 3.64 & 8.9   & 3.6 & 87       \\
\hline \hline
\end{tabular} 
\end{center}
\end{table}

The magnitude of $g_{_V} (G_{_V})$ has not been determined well:
 Recent studies  of the PNJL model applied to the QCD phase diagram suggest that 
$g_{_V}$  may be comparable to or larger than  $G_{_S}$ \cite{Bratovic,Lourenco}, so that
we change its magnitude in the following range, 
\begin{eqnarray} 
  0 \le \frac{g_{_V}}{G_{_S}} \le 1.5 .
  \label{eq:GV}
\end{eqnarray}

In \S 5 and \S 6, we will show our results mainly for 
the HK parameter set with the vector interaction of the $g_{_V}$ type.
 At the end of \S 5, we discuss how the results change in  other cases.
The Q-EOS with strangeness is obtained from the above model under two conditions: 
(i) the charge neutrality among quarks and leptons, i.e.
 $\frac{2}{3}n_u-\frac{1}{3}n_d-\frac{1}{3}n_s-n_e - n_{\mu}=0$,
 and (ii) the $\beta$-equilibrium among quarks and leptons, i.e.
  $\mu_d=\mu_s=\mu_u+\mu_e$ and $\mu_e=\mu_{\mu}$.

\begin{figure}[!t]
\begin{tabular}{p{0.48\textwidth}p{0.04\textwidth}p{0.48\textwidth}}
\centering
\includegraphics*[width=7cm,keepaspectratio,clip]{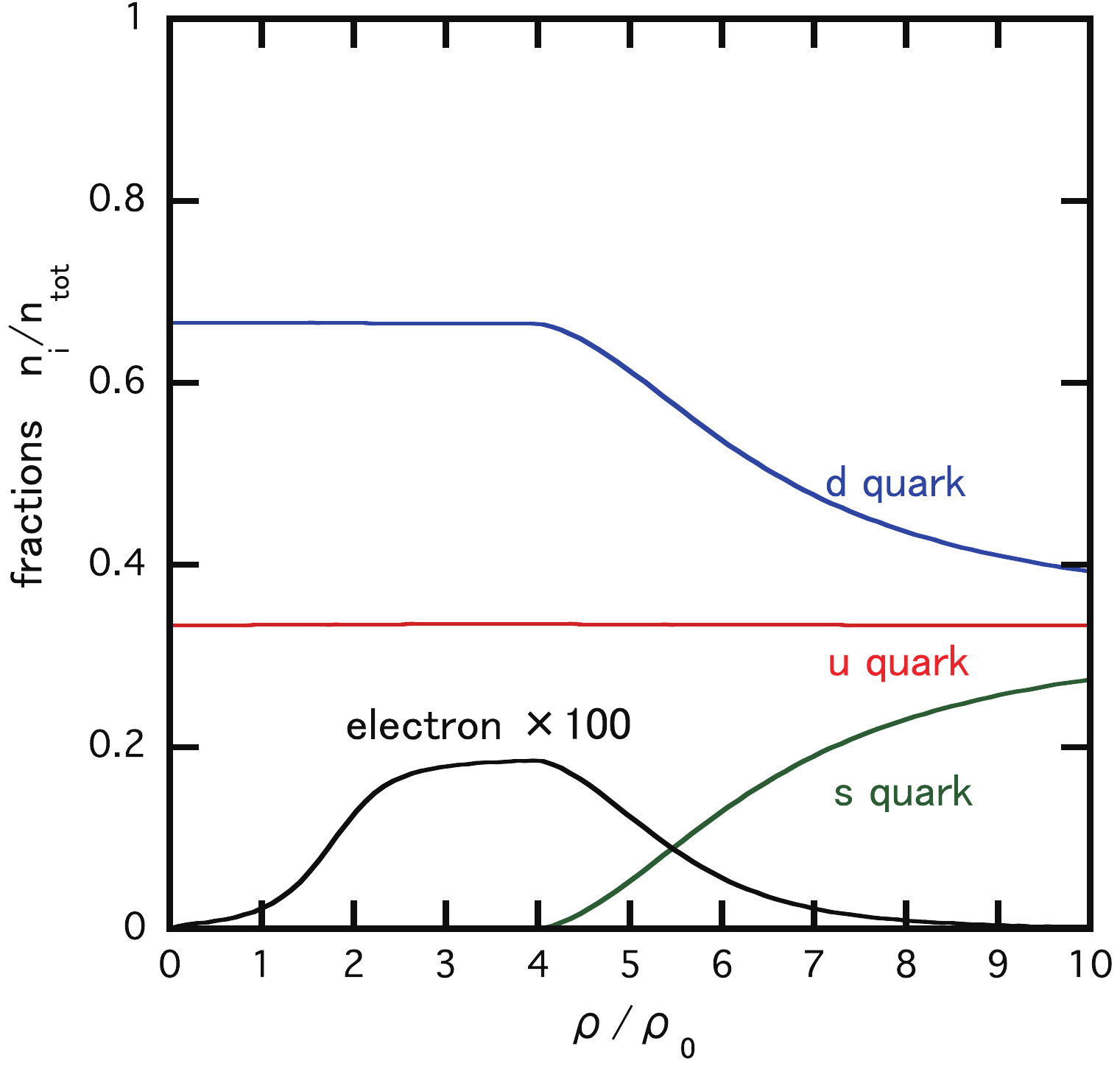}
\caption{\footnotesize{
The number fractions ($ n_{u,d,s,e}/n_{\rm tot}$ with
$n_{\rm tot}=n_u+n_d+n_s=3 \rho$) as a function of the baryon density $\rho$.
Solid red line: The fraction of u quark.  Solid blue line: The fraction of d quark. Solid green line: The fraction of s quark. Solid black line: The fraction of electron $\times$ 100. Muon does not appear due to the emergence of s quarks.}}
      \label{quark-fraction}
&&
\centering
\includegraphics*[width=7cm,keepaspectratio,clip]{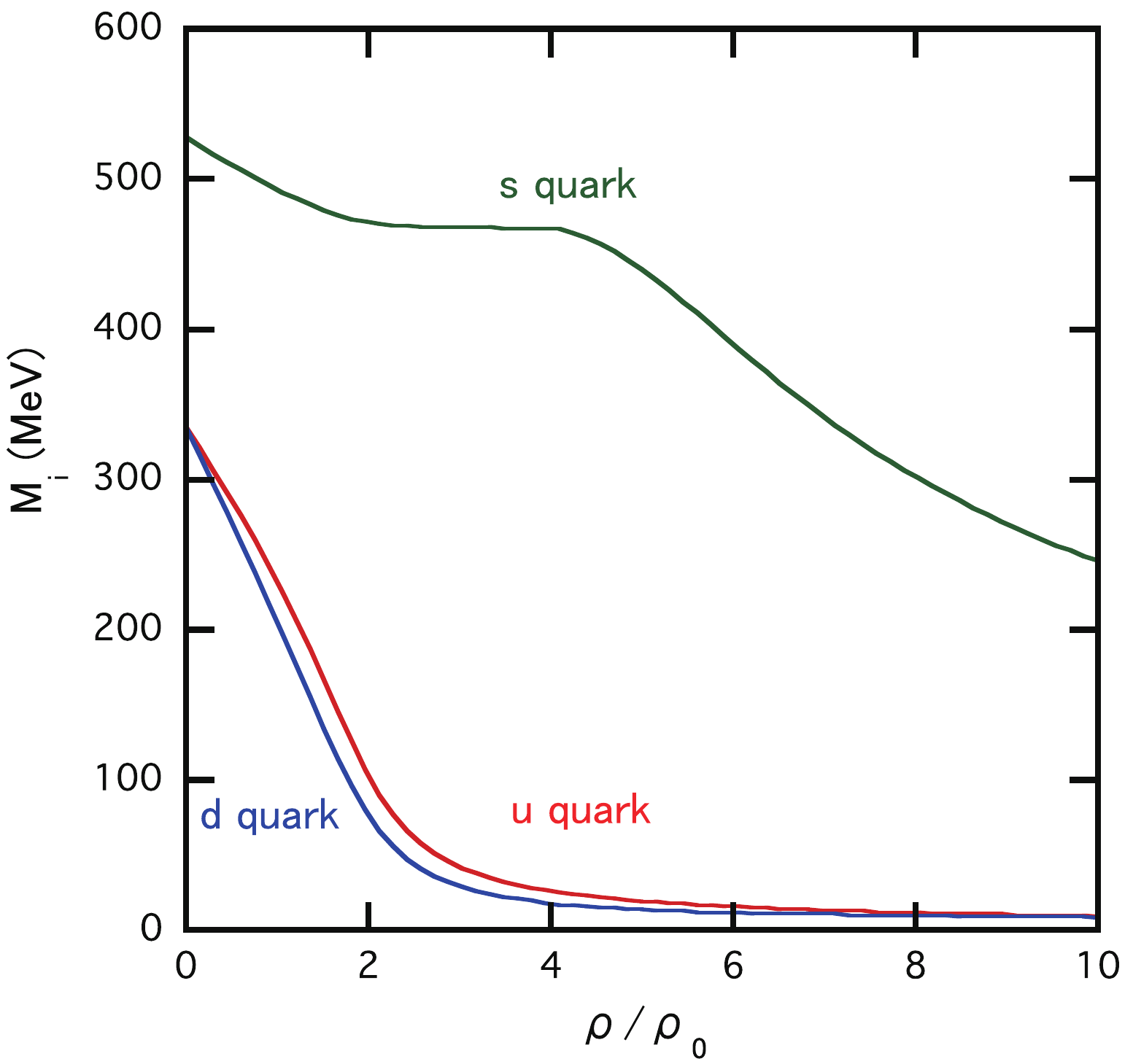}
\caption{\footnotesize{
The constituent quark masses ($M_i$) as a function of $\rho$.
Colors on each line are the same with those in Fig.\ref{quark-fraction}.      
} }
      \label{quark-mass}
\end{tabular}
\end{figure}

 In Fig.\ref{quark-fraction}, the number fractions ($ n_{u,d,s,e}/n_{\rm tot}$ with
 $n_{\rm tot}=n_u+n_d+n_s=3 \rho$) as a function of the baryon density $\rho$ are plotted. 
  Also, in Fig.\ref{quark-mass}, 
   the constituent quark masses ($M_i$) as a function of $\rho$ are plotted.
 The HK parameter set with the $g_{_V}$-type interaction are used in both figures.
  The flavor-independent $g_{_V}$-type interaction leads to a
   pressure in Eq.(\ref{eq:NJL-pressure}) depending only on  $\mu_i^{\rm eff}$. 
 Then, 
 the number fractions and the quark masses as a function of $\rho$ do
 not depend on  $g_{_V}$.
   
 At low baryon densities below a threshold density $\rho_{\rm th} \simeq 4\rho_0$, the system is composed of only $u,d$ and $e$
  with $n_d \sim 2 n_u$ due to charge neutrality and $\beta$-equilibrium (Fig.\ref{quark-fraction}).
 In this region, the  strong interaction among quarks (mainly the $G_{_S}$-term in the NJL model)
 drives
  the partial restoration of chiral symmetry and hence a rapid decrease of the constituent
  masses $M_{u,d}$ (Fig.\ref{quark-mass}).  Due to the coupling between different flavors
  through the $G_{_D}$-term, the strange quark mass $M_s$ in the Dirac sea
  is also affected slightly.

   When the baryon density exceeds $\rho_{\rm th}$, 
   the chemical potential of the strange quark $\mu_s$ becomes larger than the 
    strange quark mass ($\mu_s > M_s$), so that
    the system starts to have the strangeness degree of freedom. Since the strange quark
    is negatively charged, the electrons start to disappear from the system 
     and the $d$ quark fraction gets decreased at the same time (Fig.\ref{quark-fraction}).
    We note that the system does not have the muon, because
    the electron chemical potential is smaller than $m_{\mu}$=106MeV at all densities.
    In the high density limit, system approaches to the flavor symmetric $u,d,s$ matter
   without leptons. Once the $s$-quark appears in the system, $M_s$ is also 
    suppressed mainly due to the $G_{_S}$-term (Fig.\ref{quark-mass}).
   The strangeness threshold $\rho_{\rm th}$
    does not depend on $g_{_V}$  as already mentioned, while it depends 
    on the  NJL parameter sets in Table 2;  
      $\rho_{\rm th}/\rho_0=4.0 , 3.9$ and $3.0$ for HK, RKH and LKW, respectively.

\begin{figure}[!t]
  \begin{center}
      \centerline{\includegraphics[scale=0.55]{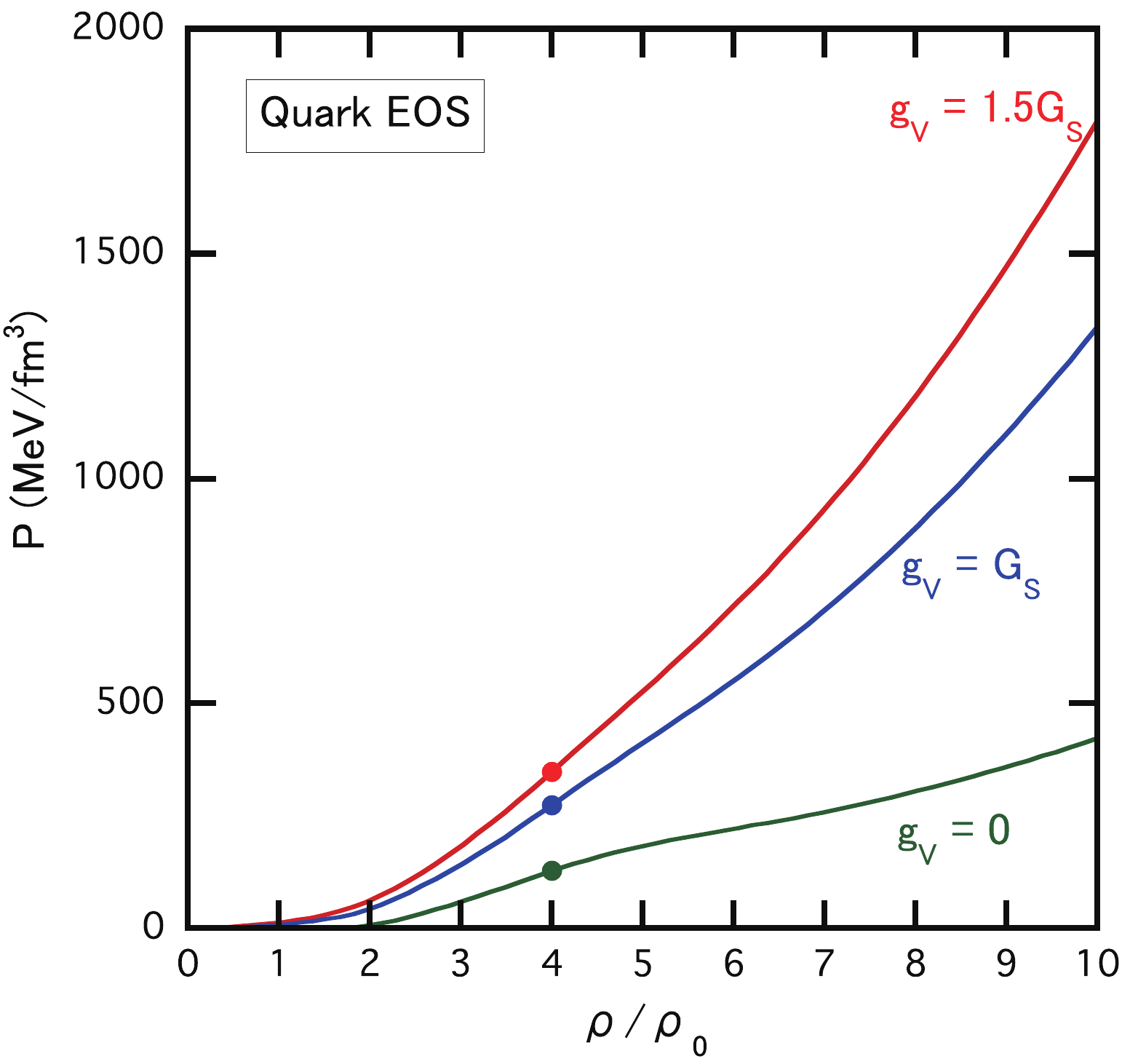}}
\end{center}
\caption{\footnotesize{ Pressure ($P$) as a function of baryon density 
$\rho$ in a pure quark matter for the  HK parameter set with
$g_{_V}/G_{_S}=0, 1.0, 1.5$.  The filled circles denote the 
 onset of the strangeness.
}}
\label{quark-pressure}
\end{figure}

In Fig.\ref{quark-pressure}, we plot the pressure ($P(\rho)$ with 
a normalization $P(0)=0$)  of the strongly interacting
quark matter for the HK parameter set with  different values of the
vector coupling ($g_{_V}/G_{_S}=0, 1.0, 1.5$ according to Eq.(\ref{eq:GV})).
Due to the universal repulsion  of the $g_{_V}$-type 
 vector interaction, the Q-EOS becomes stiffer as $g_{_V}$ increases.
 As mentioned already, the onset density of the strangeness (marked by the 
  filled circles) does not depend on $g_{_V}$.
 We note here that the present Q-EOS has a first-order phase transition
  below $2 \rho_0$ for $g_{_V} < 0.3 G_{_S}$. However, it does not
  affect the final results of the present paper, since 
  such a low density region is dominated by the hadronic EOS
  in our  hadron-quark crossover approach to be discussed in \S 4.

\section{Hadron-Quark crossover}
\label{sec:HQC}

As discussed in \S 1,  treating the point-like hadron as
an independent degree of freedom loses its validity as the
 baryon density approaches to the percolation region.
 In other words, the system cannot be described
neither by an extrapolation of the hadronic EOS
from the low-density side nor by an extrapolation of the quark EOS
from the high-density side.
Under such situation, it does not make much sense to
 apply the Gibbs criterion of two phases I and II,
 $P_{\rm I}(T_c,\mu_c)=P_{\rm II}(T_c,\mu_c)$ since $P_{\rm I}$ and $P_{\rm II}$ are not 
 reliable in the transition region.

 Since the first principle QCD calculation  at high baryon density
  is not available and effective models at finite baryon density 
  with proper treatment of the confinement phenomena  do not exist
   at present, we will consider a phenomenological ``interpolation" between
  the H-EOS and Q-EOS as a first step. Such an interpolation is certainly 
  not unique: Here we consider two simplest possibilities,
   $P$-interpolation and $\varepsilon$-interpolation as described below.
  
\begin{itemize}
\item $P$-interpolation as a function of baryon density 
\begin{eqnarray}
P(\rho) &=&P_H(\rho) f_-(\rho)+P_Q(\rho) f_+ (\rho), 
\label{eq:HQ-EOS-0} \\
f_{\pm}(\rho) &=& \frac{1}{2} \left( 1\pm \mathrm{tanh}\left( \frac{\rho-\bar{\rho}}{\Gamma} \right) \right),
\label{eq:HQ-EOS}
\end{eqnarray}
where $P_{H}$ and $P_Q$ are the pressure in the  hadronic matter and that in  
the  quark matter, respectively.  The interpolating
  function $f_{\pm}$ similar to ours has been previously considered at finite temperature 
 in \cite{Asakawa-Hatsuda,Blaizot-1,Blaizot-2}.
The window $ \bar{\rho} - \Gamma \ltsim \rho \ltsim  \bar{\rho} + \Gamma$
 characterizes the crossover region in which 
both hadrons and quarks are strongly interacting, so that neither pure hadronic EOS nor pure quark EOS are reliable.  The percolation picture illustrated in Fig.1 is best implemented 
 by the interpolation in terms of the baryon density $\rho$ instead of the baryon chemical
  potential.    One should not confuse Eq.(\ref{eq:HQ-EOS}) with the 
  pressure in the mixed phase associated with the first-order phase transition in which
   $f_{\pm}$ is considered to the volume fraction of each phase.  In our crossover picture,
   the system is always uniform and  $f_{-}$ ($f_{+}$) 
  should be interpreted as
   the degree of reliability of H-EOS (Q-EOS)  at given baryon density.
 
 To calculate the energy density $\varepsilon$ as a function of $\rho$ 
 in thermodynamically consistent way, we integrate the thermodynamical relation,  
 $P=\rho^2 {\partial(\varepsilon/\rho)}/{\partial \rho}$ and obtain 
\begin{eqnarray}
  \varepsilon (\rho)
  &=&\varepsilon_H (\rho) f_- (\rho)+\varepsilon_Q (\rho) f_+(\rho) 
  +\Delta \varepsilon  \\
\Delta \varepsilon&=&\rho \int^{\rho}_{\bar{\rho}}(\varepsilon_H(\rho')
-\varepsilon_Q(\rho'))\frac{g(\rho')}{\rho'}d\rho'  
\end{eqnarray}
with $g(\rho)=\frac{2}{\Gamma}(e^X+e^{-X})^{-2}$ and $X=(\rho-\bar{\rho})/{\Gamma}$.
Here $\varepsilon_H$ ($\varepsilon_Q$) is the energy density obtained
 from H-EOS (Q-EOS). 
  $\Delta \varepsilon$ is an extra term which guarantees the thermodynamic 
  consistency. Note that the energy per baryon from the 
  extra term $\Delta \varepsilon/\rho $, which receives main contribution from the 
  crossover region, is  finite even in the  high-density limit.

\item  $\varepsilon$-interpolation as a function of baryon density 
\begin{eqnarray}
\varepsilon(\rho) =\varepsilon_H(\rho) f_-(\rho)+\varepsilon_Q(\rho) f_+ (\rho).
\label{eq:e-interpolation}
\end{eqnarray}
Other thermodynamic quantities are obtained through the thermodynamic relation;
\begin{eqnarray}
P(\rho)% &=&\rho^2 \frac{\partial(\varepsilon/\rho)}{\partial \rho} \nonumber \\
&=&P_H (\rho) f_- (\rho)+P_Q (\rho) f_+(\rho) +\Delta P  \\
\Delta P&=&\rho (\varepsilon_Q(\rho)-\varepsilon_H(\rho))g(\rho),  
\label{eq:Delta-P}
\end{eqnarray} 
and $\mu=(\varepsilon+P)/{\rho}$. Here
 $\Delta P$ is an extra term which guarantees the thermodynamic consistency;
  it is a localized function in the crossover region and  obeys the 
 property, $\Delta P(0)=\Delta P(\infty)=0$.  
\end{itemize}

\section{Neutron star properties with $P$-interpolation}

\subsection{Interpolated EOS}

\begin{figure}[!b]
\begin{tabular}{p{0.48\textwidth}p{0.04\textwidth}p{0.48\textwidth}}
\centering
\includegraphics*[width=7cm,keepaspectratio,clip]{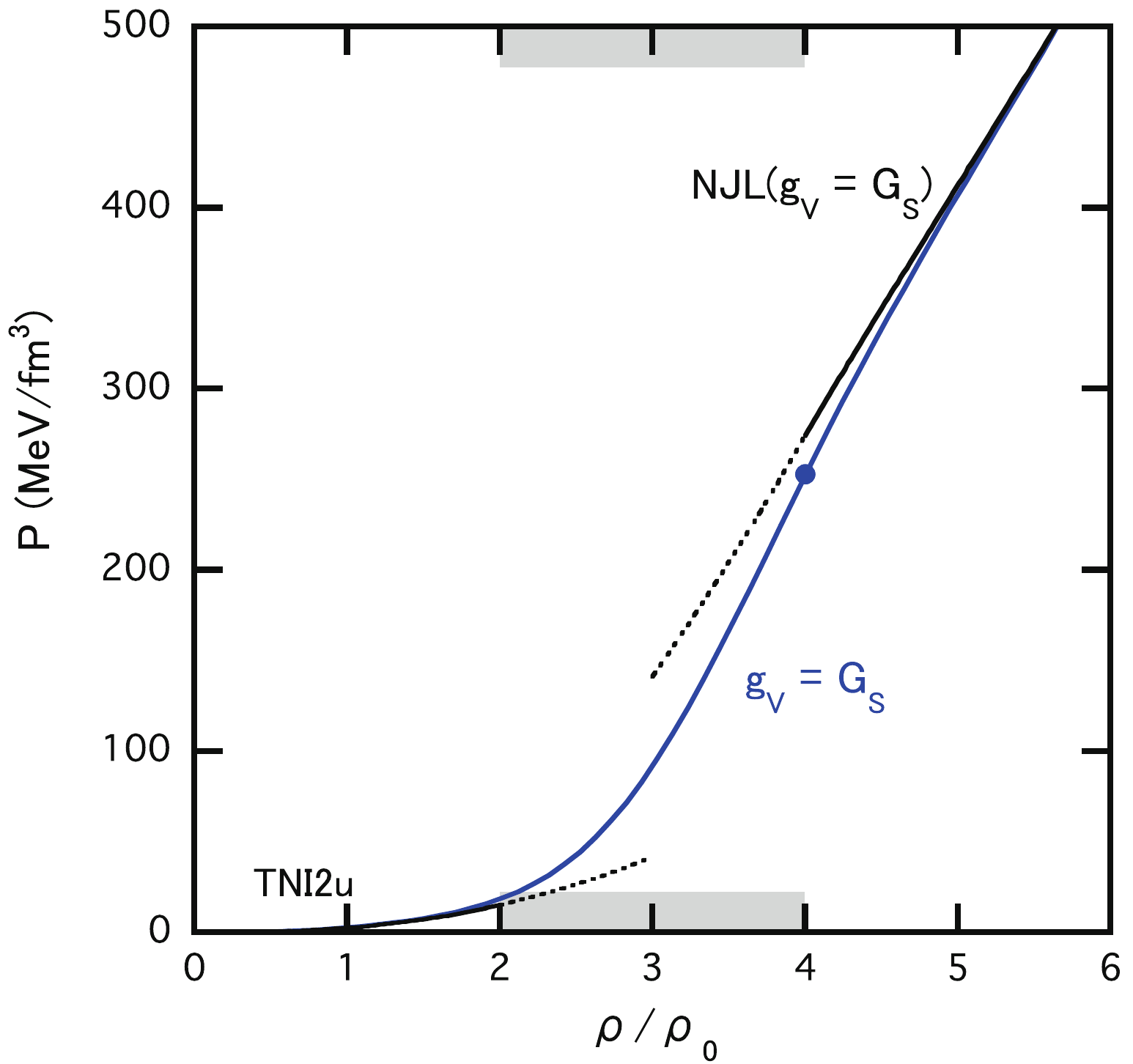}
\caption{\footnotesize{The interpolated pressure between TNI2u H-EOS and NJL Q-EOS with $g_{_V}=G_{_S}$ for $(\bar{\rho}, \Gamma) = (3\rho_0, \rho_0)$. 
Pressure is illustrated by a blue line.
The filled circle denotes the threshold density of strangeness. } }
      \label{interpolate-p-tni2u}
&&
\centering
\includegraphics*[width=7cm,keepaspectratio,clip]{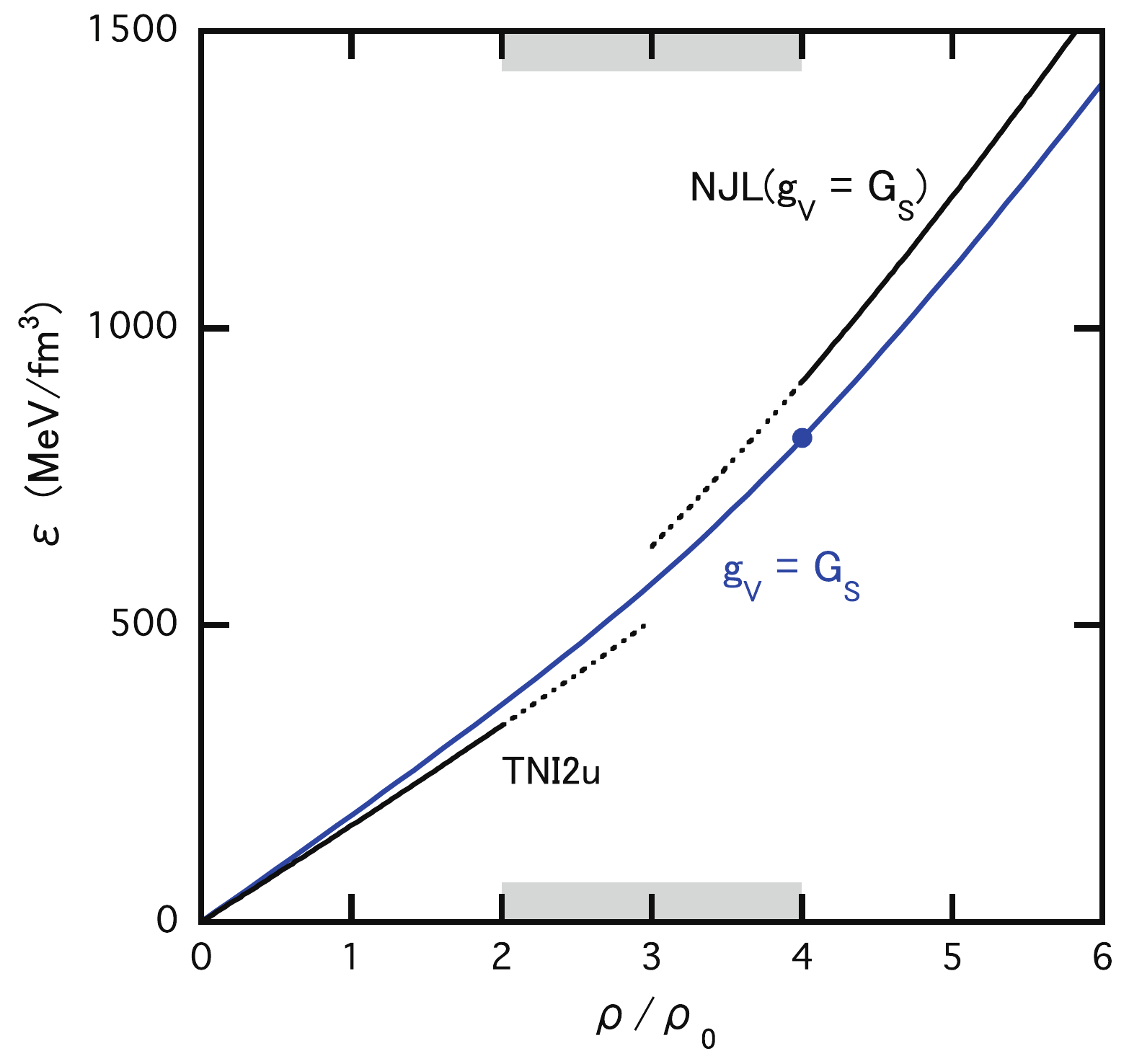}
 \caption{\footnotesize{
The energy density obtained from the 
 interpolated pressure in Fig.\ref{interpolate-p-tni2u}. 
 Energy density is illustrated by a blue line.
The filled circle denotes the threshold density of strangeness.
}}
      \label{interpolate-e-tni2u}
\end{tabular}
\end{figure}

  In the present section we consider the 
  case of $P$-interpolation. The case of
 $\varepsilon$-interpolation will be discussed in \S 6.  
    We note that the crossover window in both interpolations 
   should  satisfy the following physical conditions: 
(i) The system is always thermodynamically stable $dP/d\rho >0$, and
(ii) the normal nuclear matter is well described by the H-EOS
 so that   $ \bar{\rho} -2 \Gamma > \rho_0 $ is satisfied.
 
\begin{figure}[!t]
\begin{tabular}{p{0.48\textwidth}p{0.04\textwidth}p{0.48\textwidth}}
\centering
\includegraphics*[width=7cm,keepaspectratio,clip]{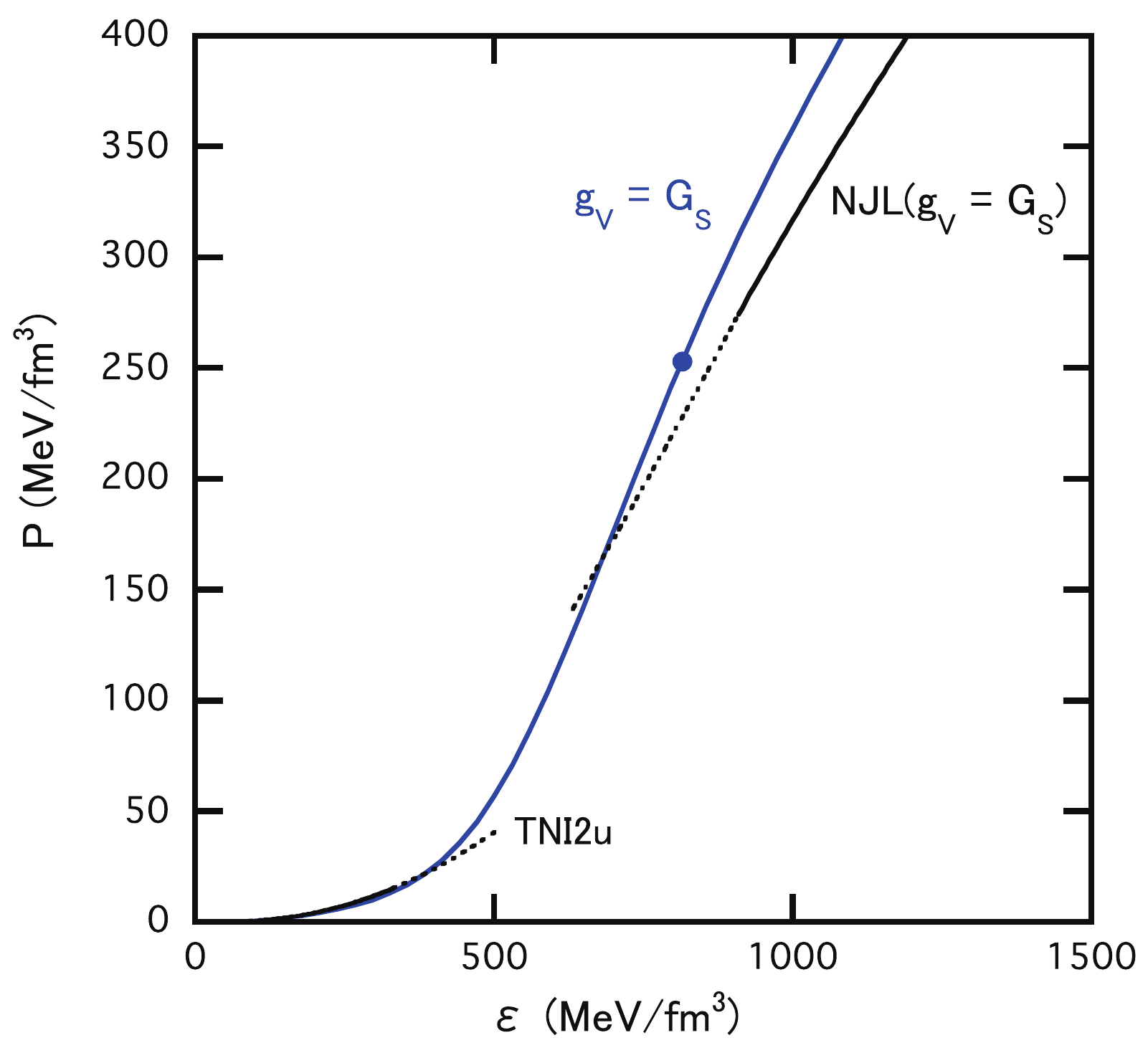}
\caption{\footnotesize{The relation between interpolated pressure and the  energy density.  
The parameters are same as Fig. \ref{interpolate-p-tni2u}.
The filled circle denotes the threshold density of strangeness. } }
      \label{interpolate-pe-tni2u}
&&
\centering
\includegraphics*[width=7cm,keepaspectratio,clip]{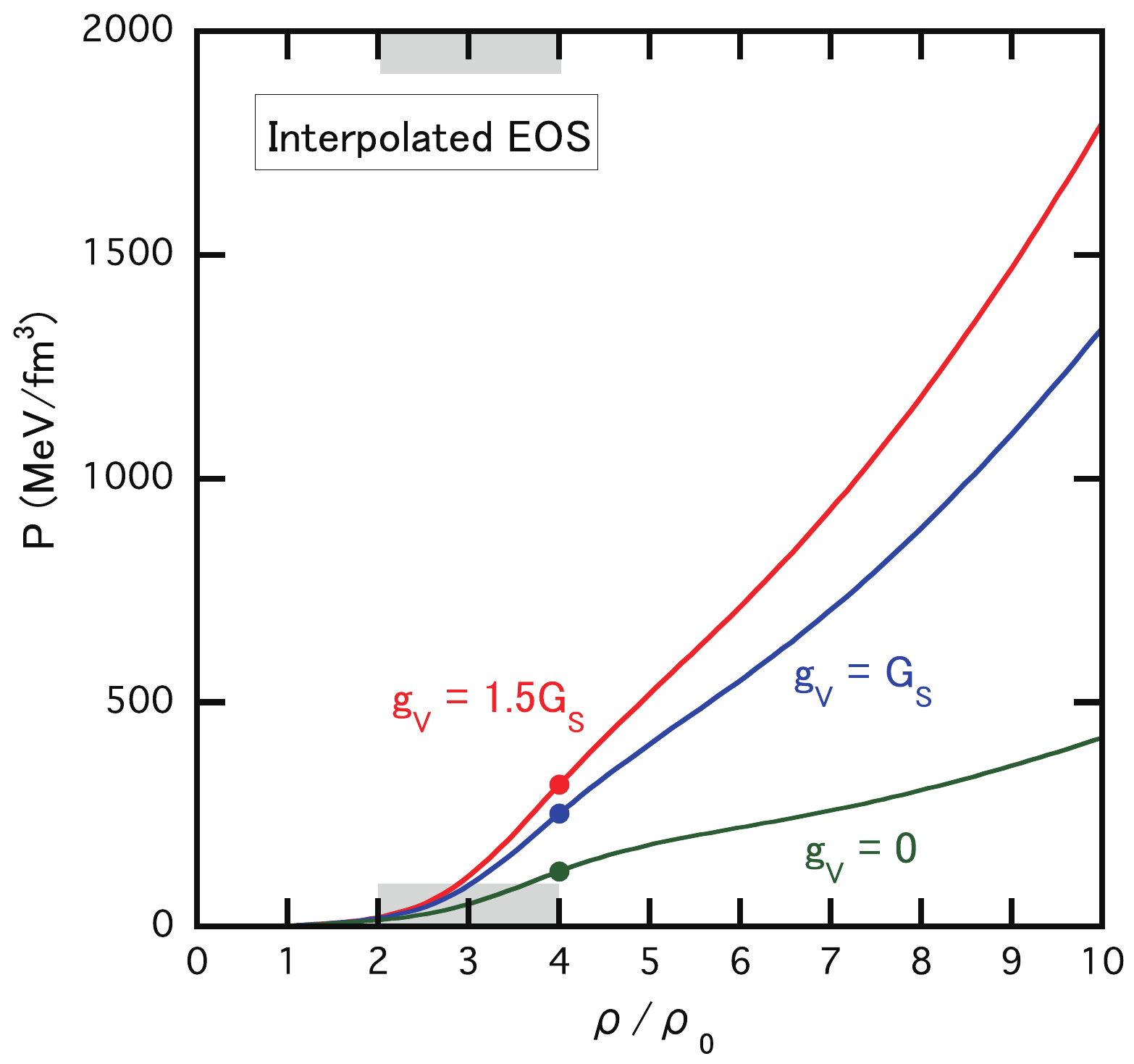}
 \caption{\footnotesize{
Interpolated pressure ($P$) as a function of baryon density $\rho$ for the case $(\bar{\rho}, \Gamma) = (3\rho_0, \rho_0)$
 with $g_{_V}/G_{_S}=0, 1.0, 1.5$.
}}
      \label{p-tni2u}
\end{tabular}
\end{figure}

 Shown in Fig.\ref{interpolate-p-tni2u}, Fig.\ref{interpolate-e-tni2u} and Fig.\ref{interpolate-pe-tni2u} are  examples  of the $P$-interpolation between 
  TNI2u for H-EOS and NJL with $g_{_V}=G_{_S}$ for Q-EOS
   according to Eq.(\ref{eq:HQ-EOS}).
  The crossover window is chosen to be  $(\bar{\rho}, \Gamma) = (3\rho_0, \rho_0)$
   and is shown by the shaded area on the horizontal axis.
  An important lessen one can learn from  Fig.\ref{interpolate-p-tni2u} is that
  the H-EOS (Q-EOS) is  nothing more than  the 
   asymptotic form of the ``true" $P(\rho)$ around $\rho=0$ ($\rho=\infty$). Therefore,  
  naive extrapolation of H-EOS and Q-EOS beyond their applicability 
    would miss essential physics.

 In Fig.\ref{p-tni2u}, we plot the interpolated EOS using TNI2u and NJL
 for different values
  of $g_{_V}$ in a wide range of baryon density. The filled circles 
  denote the onset of strangeness degrees of freedom, either hyperons or strange quarks.

\subsection{Mass-radius relation}

We now solve the following Tolman-Oppenheimer-Volkov (TOV) equation 
 to obtain $M$-$R$ relationship by using 
 the  EOSs with and without the hadron-quark crossover:
\begin{eqnarray}
\frac{dP}{dr}&=&-\frac{G}{r^2}\left(M(r)+4\pi Pr^3\right)\left(\varepsilon +P\right)\left(1-2GM(r)/r\right)^{-1} , \nonumber \\
M(r)&=&\int^r_0 4\pi r'^2 \varepsilon(r') dr' ,
\end{eqnarray}
where we have assumed the spherical symmetry with $r$ being the 
 radial distance from the center of the star.

\begin{figure}[!h]
  \begin{center}
      \centerline{\includegraphics[scale=0.6]
                                     {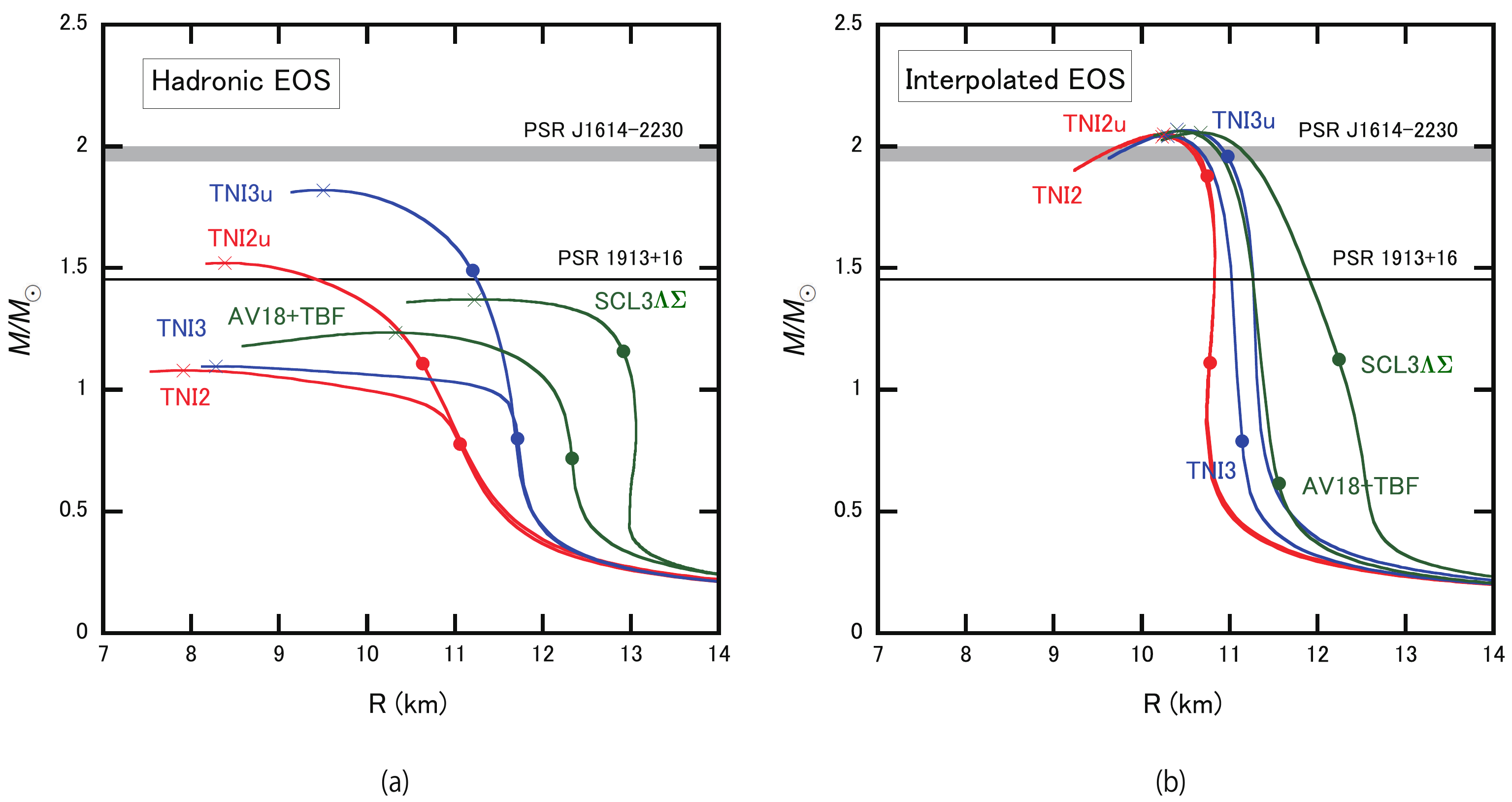}}
\end{center}
      \caption{\footnotesize{
$M-R$ relationships.
(a)$M-R$ relationships with various H-EOS including hyperons. Solid red lines: TNI2u (universal 3-body force  with $\kappa=250$MeV) 
 and TNI2 (3-nucleon force with $\kappa=250$MeV).  
 Solid blue lines: TNI3u (universal 3-body force with $\kappa=300$MeV) 
 and TNI3 (3-nucleon force with $\kappa=300$MeV) \cite{Nishizaki_2001,Nishizaki_2002}.
 Solid green lines: AV18+TBF (G-matrix approach with hyperons) \cite{Baldo} 
 and SCL3$\Lambda \Sigma$ (Relativistic mean field model with a chiral SU(3) symmetry)
  \cite{Tsubakihara}. 
 The gray band denotes $M=(1.97\pm0.04)M_{\odot}$ for PSR J1614-2230.
 The solid black line denotes $M=1.44M_{\odot}$ for PSR 1913+16.
(b) $M$-$R$ relationship with the EOS interpolated between H-EOS in (a) and Q-EOS with the HK parameter set and $g_{_V}=G_{_S}$, by the window parameters $(\bar{\rho},\Gamma)=(3\rho_0,\rho_0)$.
Colors on each line are the same with those in (a).  
}}
      \label{various EOS}
 \end{figure}

In Fig. \ref{various EOS}(a), we show the $M$-$R$ relationship for various H-EOSs with hyperons
 whose  onset is denoted by the filled circles.
 The crosses denote the points where maximum masses are realized:
  In all cases,   
 $M_{{\rm max}}$  does not reach $2M_{\odot}$ due to the 
  softening of EOS by the hyperon mixture.  
 
 In Fig. \ref{various EOS}(b), we show the $M$-$R$ relationship with
  the EOS interpolated between H-EOS and Q-EOS:  For the H-EOS, we consider the same
  EOSs as shown in Fig. \ref{various EOS}(a), while for the Q-EOS, we 
 adopt  the  HK-parameter set with $g_{_V}=G_{_S}$ as a typical example.
 The crossover window  is fixed to be $(\bar{\rho},\Gamma)=(3\rho_0,\rho_0)$.
  Cases for different  parameters in Q-EOS 
 as well as for different window parameters are discussed in the next subsection. 

  The red lines in Fig. \ref{various EOS}(b) 
  correspond to the cases with TNI2u and TNI2, the blue lines correspond
 to TNI3u and TNI3, and the green lines correspond to SCL3$\Lambda \Sigma$ and AV18+TBF.
 The onset of strangeness and the maximum mass are denoted by the filled circles and the crosses, respectively.
 Irrespective of the H-EOSs, the interpolated EOS can sustain hybrid star with 
  $M_{{\rm max}} > 2M_{\odot}$: A smooth crossover around $\rho\sim 3 \rho_0$
   and the stiff Q-EOS due to 
  repulsive vector interaction are two fundamental reasons behind this fact. 
  Also, we  note that the radius of the hybrid star with interpolated EOS
   is in a range $R= (11 \pm 1) $km for 
    $0.5 < M/M_{\odot} < 2.0$, except for the case of SCL3$\Lambda\Sigma$.
   \footnote{The reason why the case with SCL3$\Lambda\Sigma$ is different from others 
  can be easily seen from Fig. \ref{PwithY}:
  The pressure $P$ of SCL3$\Lambda \Sigma$ is nearly twice as large as that of the other EOSs
  at  $\rho=(1-2)\rho_0$. This leads to a larger $R$ of light NSs.}
   Such a narrow window of  $R$ independent of the values of $M$   
  is consistent with the phenomenological constraints on $R$ 
  based on recent observations of both transiently accreting and bursting sources \cite{Steiner,Ozel_2012}.

\begin{table}[!h]
\caption{\footnotesize{
$M_{\rm max}/M_{\odot}$ ($\rho_c/\rho_0$) for 
 different choice of H-EOS and difference stiffness of Q-EOS.
 }} 
\label{table 2}
\begin{center}
  \begin{tabular}[c]{c|c|c}  \hline \hline
   H-EOS 
 & ${g_{_V}}$=${G_{_S}}$ 
 & ${g_{_V}}$=$1.5{G_{_S}}$    \\ \hline 
 TNI2u   & 2.05 (6.1) & 2.17 (5.5)    \\ 
 TNI2    & 2.04 (6.1) & 2.16 (5.9)    \\ 
 TNI3u   & 2.07 (5.9) & 2.18 (5.4)    \\ 
 TNI3    & 2.04 (6.1) & 2.16 (5.5)    \\
Paris+TBF & 2.06 (6.1) & 2.17 (5.6)   \\
 AV18+TBF & 2.06 (6.1) & 2.17 (5.5)     \\  
 SCL3$\Lambda \Sigma$& 2.06 (5.9)& 2.17 (5.5)     \\ 
\hline \hline
\end{tabular} 
\end{center}
\end{table}

 In Table \ref{table 2}, we show the maximum mass and 
 the associated central density of the hybrid star with the interpolated EOS
  with  ${g_{_V}}$=${G_{_S}}$ and  ${g_{_V}}$=$1.5{G_{_S}}$.
  In all combinations of H-EOS and Q-EOS,    $M_{{\rm max}}$ exceeds $2M_{\odot}$ 
  with the central density, $\rho_c=(5.4-6.1) \rho_0$.

\begin{figure}[!h]
  \begin{center}
      \centerline{\includegraphics[scale=0.7]
                                     {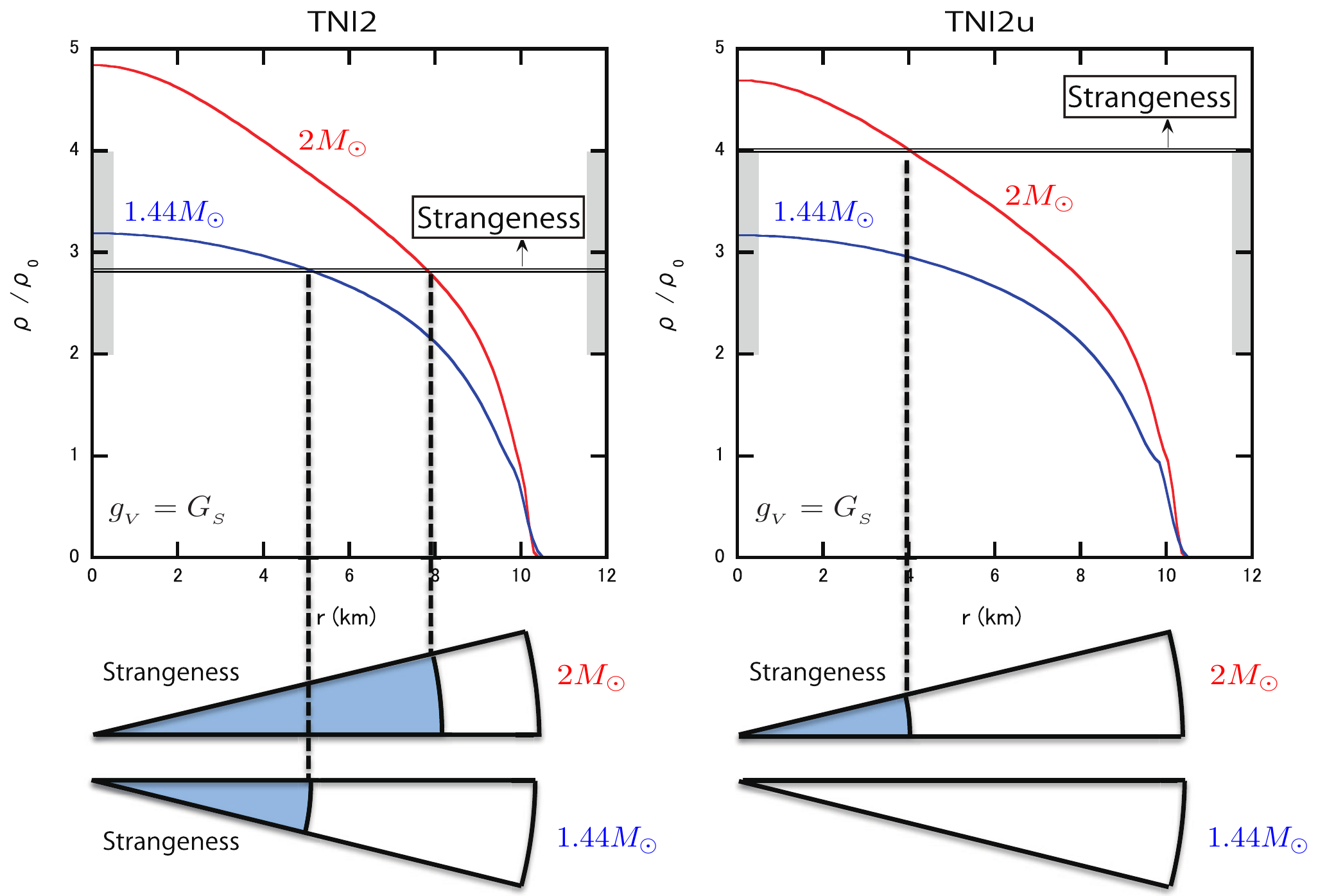}}
\end{center}
      \caption{\footnotesize{
Density-profiles $\rho(r)$ with $r$ being 
the distance from the center for the $2.0M_{\odot}$ star (red line) 
and $1.44M_{\odot}$ star (blue line).  
In the left upper panel TNI2 H-EOS and Q-EOS with $g_{_V}=G_{_S}$ 
and the HK parameter set are used in the interpolation with
the window parameters $(\bar{\rho},\Gamma)=(3\rho_0,\rho_0)$, while in the right
upper panel TNI2u H-EOS and Q-EOS above are used.
  Double line shows the density above which the strangeness appears.
  Lower illustrations show the internal structure. Only the shaded 
  regions contain strangeness degrees of freedom.
}}
      \label{structure}
 \end{figure} 
      
Let us now turn to the internal structure of the hybrid star, in particular its
strangeness content. From the location of the 
filled circles in Fig. \ref{various EOS}(b), one finds that
 the flavor-independent universal three-baryon repulsion in TNI2u and TNI3u
 increases the onset density of the strangeness inside the hybrid star.
 This can be seen more explicitly by plotting the radial profile of the 
  hybrid star:  The upper panels of Fig.\ref{structure} show the
  $\rho-r$ relationships for $2M_{\odot}$ and $1.44M_{\odot}$ hybrid stars with
 TNI2 (left) and TNI2u (right). The threshold densities of the strangeness given in
 Table \ref{property of H-EOS} are indicated by the double lines.
  In our interpolated EOSs, the above  stars turn out to have
 almost the same radius.  The lower illustrations of Fig.\ref{structure}
  show the cross sections of the corresponding hybrid stars. 
 
  These figures imply that, even if the mass and the radius are the same,
  the strangeness content of the hybrid stars can be quite different. 
  This point is of particular interest for the cooling problem of NSs. 
 As is well known, NSs with a $Y$-mixed core undergo an extremely rapid cooling due to the efficient $\nu$-emission processes called ``hyperon direct URCA" ($Y$-Durca, e.g., $\Lambda\rightarrow p+e^-+\bar{\nu}_e$, $p+e^- \rightarrow \Lambda +\nu_e$) and are cooled very rapidly below the detection limit of thermal X-ray. Therefore, for the NSs consisting of pure hadronic components with $Y$, only the very light-mass NSs ($M<(1.0-1.2)M_{\odot}$,  as in Fig. \ref{M-rho-hadron}) can escape from $Y$-Durca rapid cooling. This means an unlikely situation that all the NSs whose $T_s$ are observed should be 
 light-mass stars in spite of the fact that the observed mass distribution is
  centered around $(1.4-1.5)M_{\odot}$ \cite{NS-review}. On the contrary, in the case of the hybrid star with $g_{_V}=G_{_S} (1.5G_{_S})$ under consideration, NSs as heavy as up to $1.9(2.0)M_{\odot}$ can avoid this rapid cooling, allowing the $T_s$-observed NSs to be from the light-mass to heavy mass stars ($M \leq (1.9-2.0)M_{\odot}$, as in Fig. \ref{m-rho-tni2u})\footnote{ However, in the case of the hybrid star with smooth crossover, ``quark direct URCA" ($Q$-Durca) in stead of $Y$-Durca  
  may takes place in the crossover region. The effect of spin-singlet and spin-triplet
  color superconductivity on this $Q$-Durca is an interesting open question to be studied.}.

It is in order here to comment on the relationship between the maximum mass and the nuclear incompressibility $\kappa$.  From the properties of finite nuclei, 
the nuclear incompressibility $\kappa$ is estimated to be
 $(240\pm20)$MeV \cite{Shlomo}. The interpolated EOSs with TNI2 and TNI2u are consistent 
  with this empirical $\kappa$, and yet they can reach  $M_{\rm max} > 2M_{\odot}$.    
 In other words, what is important to sustain massive hybrid stars is not the 
  value of the incompressibility, but the stiffness of the EOS at and above
  $\sim 3\rho_0$.

\subsection{Dependence on Q-EOS}

\begin{figure}[!h]
  \begin{center}
      \centerline{\includegraphics[scale=0.6]
                                     {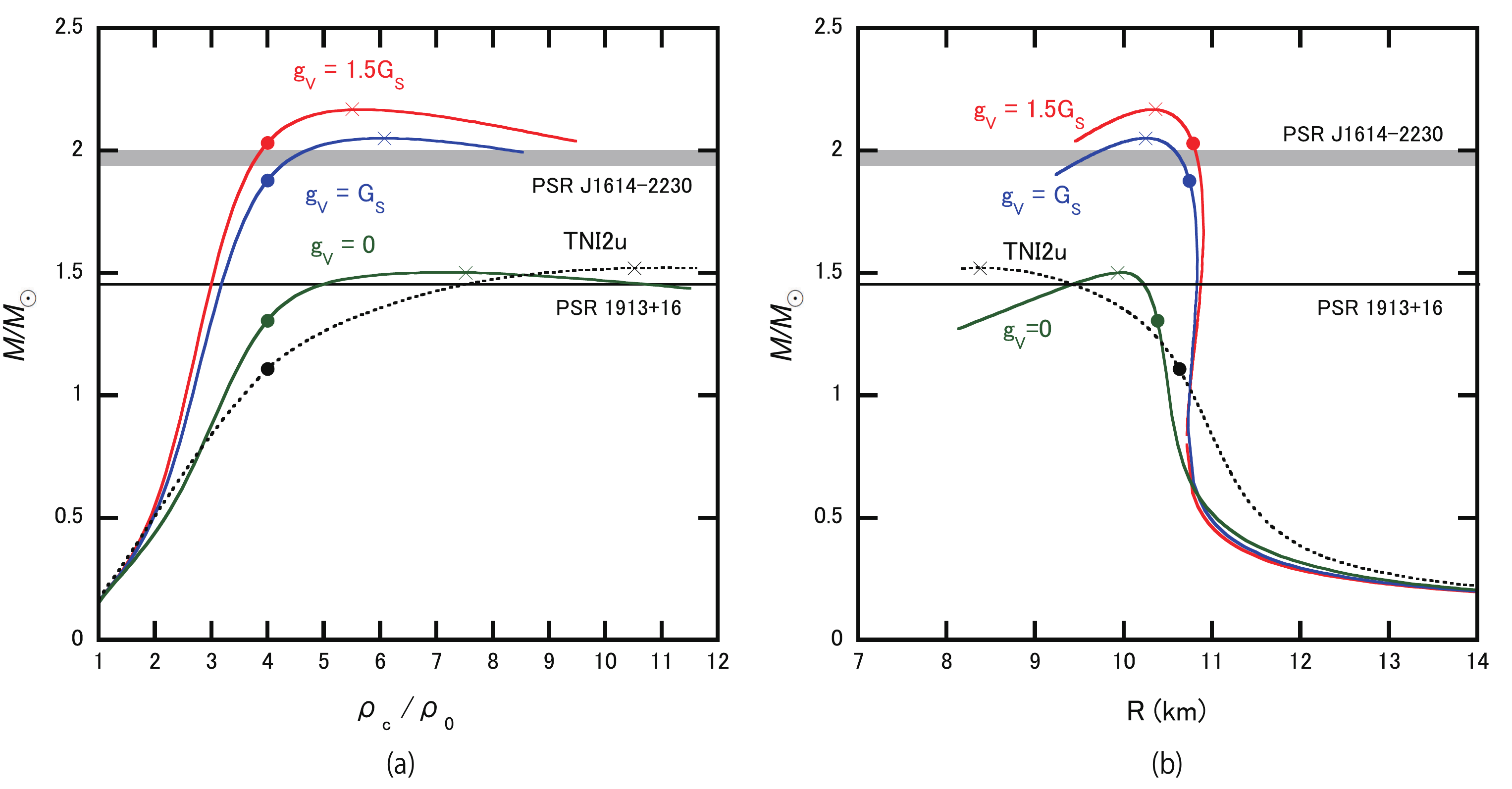}}
\end{center}
      \caption{\footnotesize{
(a) $M-\rho_c$ relationships with the interpolated EOSs.
We adopt  the  HK-parameter set for the Q-EOS with various  $g_{_V}/G_{_S}=0, 1.0, 1.5$.
 The crossover window  are fixed to be $(\bar{\rho},\Gamma)=(3\rho_0,\rho_0)$.
The cross symbols denote the points of $M_{\rm max}$, while 
the filled circles denote the points beyond which the strangeness appears.
The gray band denotes $M=(1.97\pm0.04)M_{\odot}$ for PSR J1614-2230.
 The solid black line denotes $M=1.44M_{\odot}$ for PSR 1913+16.  
(b) $M$-$R$ relationships with the interpolated EOSs.  
 }}
      \label{m-rho-tni2u}
 \end{figure} 
 
 To see how the hybrid star structure changes by the stiffness of Q-EOS,
 we plot $M-\rho_c$ relationship for $g_{_V}/G_{_S}=0, 1.0, 1.5$ with HK parameter set
 in Fig.\ref{m-rho-tni2u}(a). We take TNI2u for H-EOS and the same crossover window as in
 Fig.\ref{various EOS}.  For comparison, the $M-\rho_c$ relationship only with TNI2u   is
 plotted by the dashed line.  Fig.\ref{m-rho-tni2u}(b) shows the corresponding $M-R$ relations.
 As anticipated, $M_{\rm max}$ increases as $g_{_V}$ increases.
 In Table \ref{table 3}, we show how $M_{\rm max}$ and $\rho_c$  depend
 on the choice of $g_{_V}$ and the choice of the NJL parameter set.
 Although the parameter dependence is not entirely negligible,
  the massive hybrid star is possible for sufficiently large values of  $g_{_V}$.

\begin{table}[!h]
\caption{\footnotesize{
The values of $M_{\rm max}/M_{\odot}$ ($\rho_c/\rho_0$) for ${g_{_V}}/{G_S}=1.0, 1.5, 2.0$
 with  $(\bar{\rho},\Gamma)=(3\rho_0,\rho_0)$ and TNI2u.  The parameter sets of the 
 NJL model, HK, RKH and LKW, are given  in Table \ref{table NJL}. }} 
\label{table 3}
\begin{center}
  \begin{tabular}[c]{c|c|c|c}  \hline \hline
   Q-EOS 
 & ${g_{_V}}$=${G_S}$ 
 & ${g_{_V}}$=$1.5{G_S}$
 & ${g_{_V}}$=$2{G_S}$    \\ \hline 
 HK      & 2.05 (6.1) & 2.17 (5.5) & 2.24 (5.4)    \\ 
 RKH     & 1.99 (6.2) & 2.12 (5.8) & 2.20 (5.4)    \\ 
 LKW     & 1.72 (7.5) & 1.87 (6.7) & 1.97 (6.3)   \\ 
\hline \hline
\end{tabular} 
\end{center}
\end{table}

Finally, we consider  the flavor-dependent vector interaction proportional to 
$G_{_V}$ given in Eq.(\ref{eq-1}).
In the high density limit where $u, d, s$ quarks have equal population,
$\langle u^{\dagger}u\rangle=\langle d^{\dagger}d\rangle=\langle s^{\dagger}s\rangle$,
 the $g_{_V}$ interaction and the $G_{_V}$ interaction have the same 
 contribution to the pressure in the mean-field approximation if we
 make the identification, $G_{_V} = \frac{3}{2}g_{_V}$.
 Motivated by this relation, we show $M_{\rm max}$ and $\rho_c$ for
 $G_{_V}/G_{_S}=1.5, 2.25, 3.0$ in  Table \ref{table 4}.
 For the density relevant to the core of the hybrid stars,
 the flavor SU(3) limit is not yet achieved due to the $s$-quark mass (see Fig. \ref{quark-fraction}).
 Therefore, the EOS for the flavor-dependent repulsion
  with $G_{_V} = \frac{3}{2}g_{_V}$ is softer than the flavor independent 
  repulsion  with $g_{_V}$. This can be seen by comparing the 
  corresponding values in Table \ref{table 4} and those in Table \ref{table 3}.
 In any case, the massive hybrid star is possible for sufficiently large values of  $G_{_V}$.

\begin{table}[!h]
\caption{\footnotesize{$M_{\rm max}/M_{\odot}$  ($\rho_c/\rho_0$) 
 for the HK parameter set with  the flavor-dependent repulsion  $G_{_S}$. 
  The crossover window is $(\bar{\rho},\Gamma)=(3\rho_0,\rho_0)$ and  the hadronic EOS is TNI2u.}}
\label{table 4} 
\begin{center}
  \begin{tabular}[c]{c|c|c}  \hline \hline
   ${G_{_V}}$=$1.5{G_{_S}}$ 
 & ${G_{_V}}$=$2.25{G_{_S}}$
 & ${G_{_V}}$=$3.0{G_{_S}}$    \\ \hline 
 1.87 (6.6)  & 1.99 (6.2) & 2.07 (5.8) \\
\hline \hline
\end{tabular} 
\end{center}
\end{table}

\subsection{Dependence on crossover window}

 In Table \ref{table 1}, we show  $M_{\rm max}$ and $\rho_c$
 for different  choice of the crossover window parameterized by 
  $\bar{\rho}$ and $\Gamma$. TNI2u and HK parameter set are adopted
   for  H-EOS and Q-EOS, respectively.
 As the crossover window becomes lower and/or wider in baryon  density,  
 the interpolated EOS becomes stiffer and $M_{\rm max}$ becomes larger. 
 To be compatible with the observed massive NS with $M=(1.97\pm 0.04)M_{\odot}$,
  the crossover needs to  occur in  $(2-4)\rho_0$.

\begin{table}[!h]
\caption{\footnotesize{$M_{\rm max}/M_{\odot}$ ($\rho_c/\rho_0$)
  under the variation of the parameters, $\bar{\rho}$ and $\Gamma$, which characterize
   the crossover window. H-EOS and Q-EOS are obtained from
   TNI2u and HK parameter set, respectively.  Columns without numbers are the 
   excluded cases corresponding to $\bar{\rho}-2\Gamma < \rho_0$ in \S \ref{sec:HQC}.}} 
\label{table 1}
\begin{center}
  \begin{tabular}[c]{c|c|c|c|c}  \hline \hline
 & \multicolumn{2}{|c|}{${\Gamma}/{\rho_0}=1$} & \multicolumn{2}{|c}{${\Gamma}/{\rho_0}=2$} \\ 
 \cline{2-3}  \cline{3-4}\cline{4-5}
   $\bar{\rho}$  
 & ${g_{_V}}$=${G_S}$ 
 & ${g_{_V}}$=$1.5{G_S}$ 
 & ${g_{_V}}$=${G_S}$ 
 & ${g_{_V}}$=$1.5{G_S}$   \\ \hline 
 $3\rho_0$   & 2.05 (6.1) & 2.17 (5.5) & $-$ & $-$    \\ 
 $4\rho_0$   & 1.89 (7.2) & 1.97 (6.8) & $-$ & $-$    \\ 
 $5\rho_0$   & 1.73 (8.2) & 1.79 (8.0) & 1.74 (8.0)& 1.80 (7.7)\\ 
 $6\rho_0$   & 1.60 (9.6) & 1.64 (9.3) & 1.62 (9.2)& 1.66 (9.0)\\
\hline \hline
\end{tabular} 
\end{center}
\end{table}

\subsection{Sound velocity of interpolated EOS}

\begin{figure}[!b]
  \begin{center}
      \centerline{\includegraphics[scale=0.55]
                                     {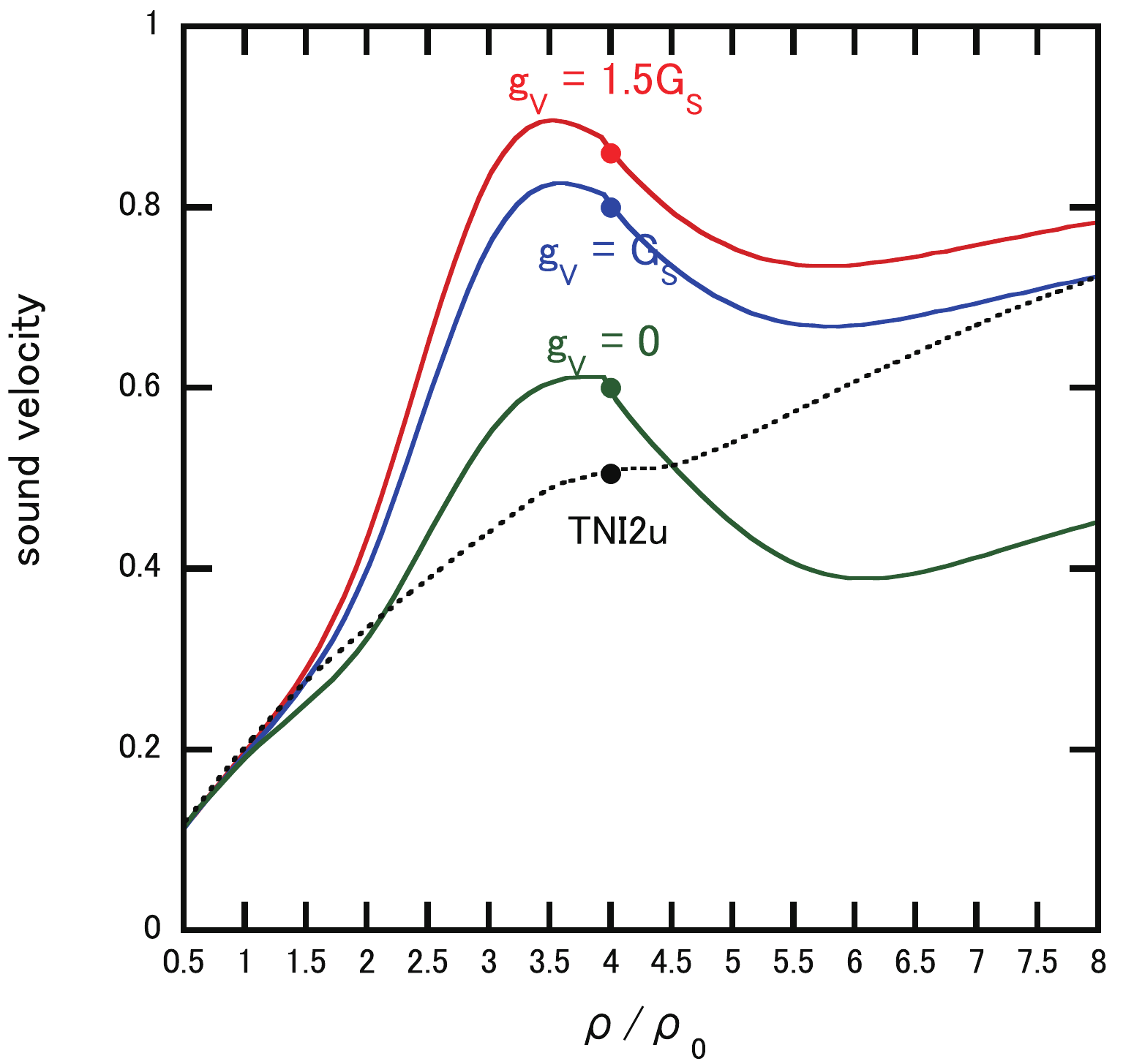}}
\end{center}
      \caption{
      \footnotesize{
Sound velocity $v_{_S}$ as a function of baryon density $\rho$.
Solid lines: $v_{_S}-\rho$ obtained from the
interpolated EOS between the H-EOS with TNI2u and the Q-EOS with $g_{_V}/G_{_S}=0, 1.0, 1.5$.
The crossover window is  $(2-4)\rho_0$. 
 Dotted line : that for  pure H-EOS with TNI2u. 
 The filled circles denote the points beyond which strangeness starts to appear.
}}

\label{sound velocity}
\end{figure}  

One of the measures to quantify the stiffness of EOS is the sound velocity 
$v_{_S}=\sqrt{dP/d\varepsilon}$.
In Fig.\ref{sound velocity}, we plot $v_{_S}$
 for our interpolated EOS with $g_{_V}/G_{_S}=0, 1.0, 1.5$
   as a function of $\rho$.
 The kinks of $v_{_S}$ at $\rho \simeq 4\rho_0$ are caused by the softening
 of EOS by the appearance of  strangeness.
 The enhancement of $v_{_S}$ of the interpolated EOS 
 relative to the pure hadronic EOS takes
  place  just at and above the crossover window.

\subsection{Stability of hybrid star}

The neutron star is  gravitationally stable
if the    average adiabatic index $\bar{\Gamma}$ satisfies
    the inequality \cite{Shapiro}: 
 \begin{eqnarray}   
\bar{\Gamma}=\frac{\int_0^R \Gamma P d^3r}{\int_0^R P d^3r} 
 > \frac{4}{3}+\lambda \frac{GM}{R}.
\label{eq:stability}
 \end{eqnarray}
Here $\Gamma=d\ln P/d \ln \varepsilon$ is the adiabatic
 index. Also, $\lambda GM/R$   
 with $\lambda$ being a numerical constant of order unity 
is a general relativistic  correction whose magnitude is 
much less than 1. 
Since $\Gamma$ of our H-EOS is about 2 at all densities
 and $\Gamma$ of our Q-EOS is larger than 4/3 due to the 
 constituent quark mass and the  repulsive vector interaction,
 Eq.(\ref{eq:stability}) is always satisfied
  and our hybrid star is gravitationally stable. 
 
\section{Neutron star properties with $\varepsilon$-interpolation}

In this section we consider the alternative 
interpolation procedure using the energy density $\varepsilon$ as a function of
 $\rho$ given in Eq.(\ref{eq:e-interpolation}).

\begin{figure}[!b]
\begin{tabular}{p{0.48\textwidth}p{0.04\textwidth}p{0.48\textwidth}}
\centering
\includegraphics*[width=7cm,keepaspectratio,clip]{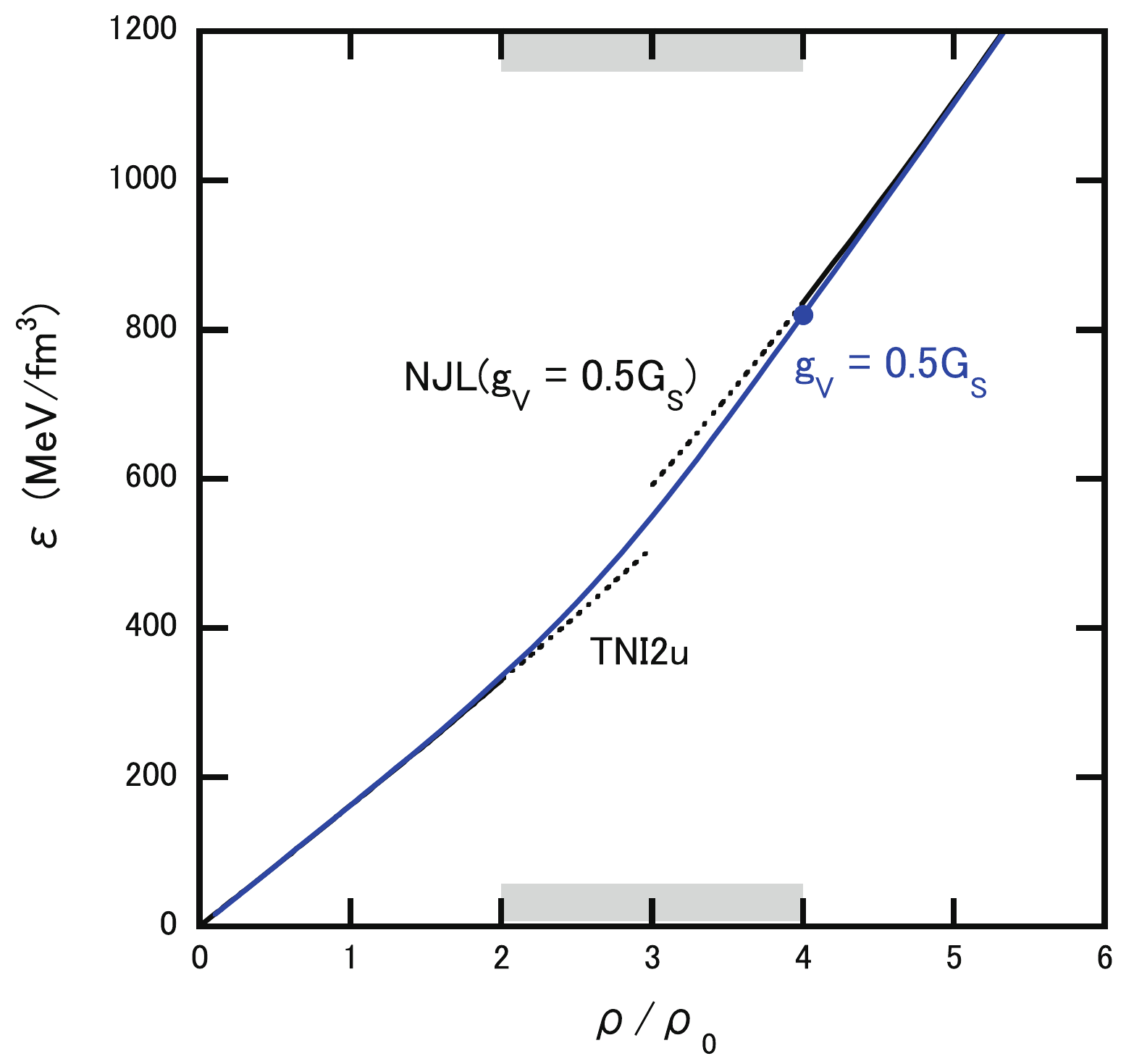}
 \caption{\footnotesize{
The interpolated energy density between TNI2u H-EOS and NJL Q-EOS with $g_{_V}=0.5G_{_S}$ for $(\bar{\rho}, \Gamma) = (3\rho_0, \rho_0)$. 
Energy density is illustrated by a blue line.
The filled circle denotes the threshold density of strangeness.}} 
  \label{interpolate-e-tni2u-energy}
&&
\centering
\includegraphics*[width=7cm,keepaspectratio,clip]{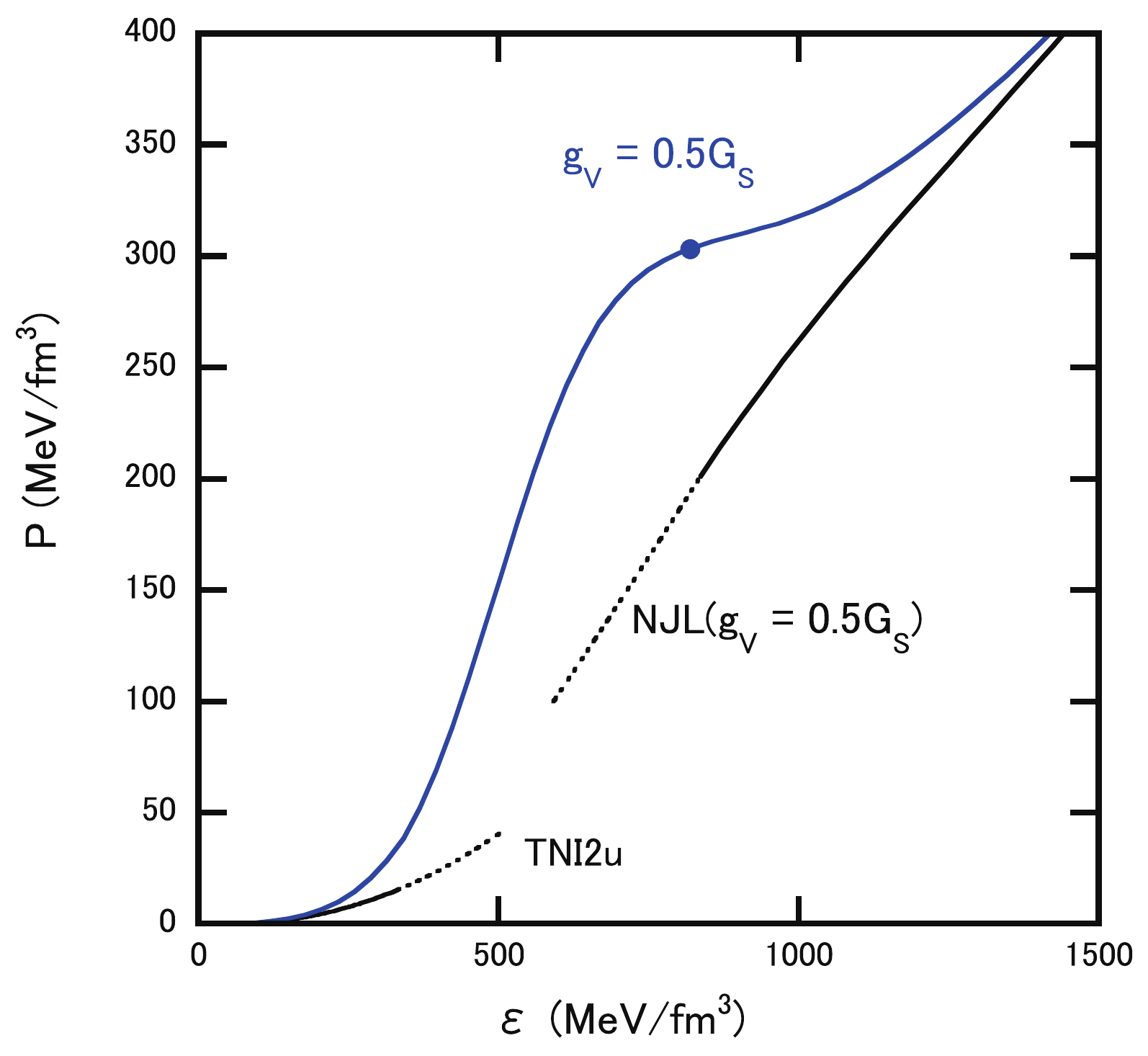}
\caption{\footnotesize{The pressure obtained from the interpolated 
 energy density in Fig.\ref{interpolate-e-tni2u-energy}. 
The pressure is illustrated by a blue line.
The filled circle denotes the threshold density of strangeness. } }
      \label{interpolate-p-tni2u-energy}
\end{tabular}
\end{figure}

\begin{figure}[!t]
\begin{tabular}{p{0.48\textwidth}p{0.04\textwidth}p{0.48\textwidth}}
\centering
\includegraphics*[width=7cm,keepaspectratio,clip]{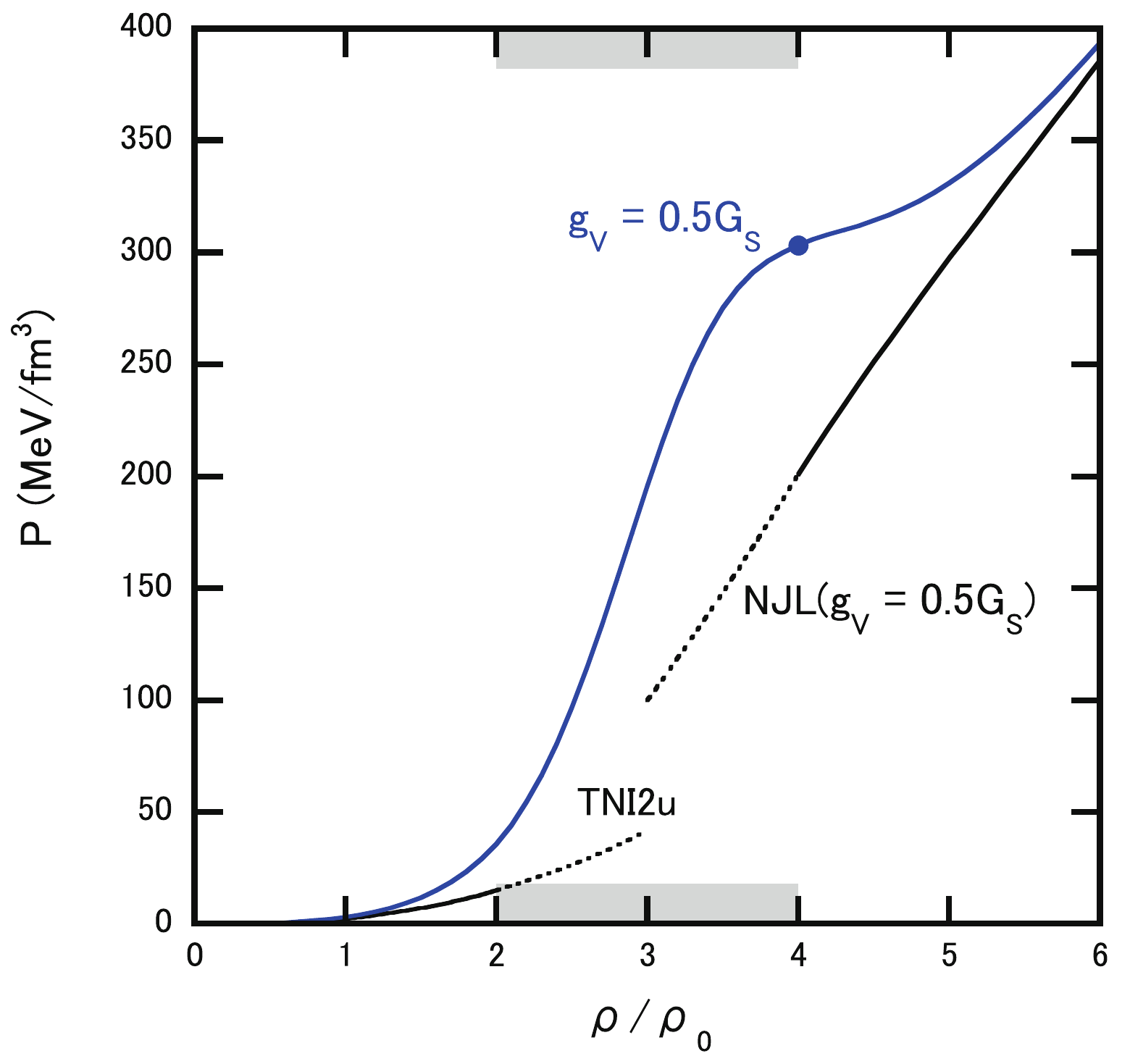}
\caption{\footnotesize{The relation between the interpolated energy density
 and the resultant pressure.  
 The parameters are same as Fig. \ref{interpolate-p-tni2u-energy}.
The filled circle denotes the threshold density of strangeness. } }
      \label{interpolate-pe-tni2u-energy}
&&
\centering
\includegraphics*[width=7cm,keepaspectratio,clip]{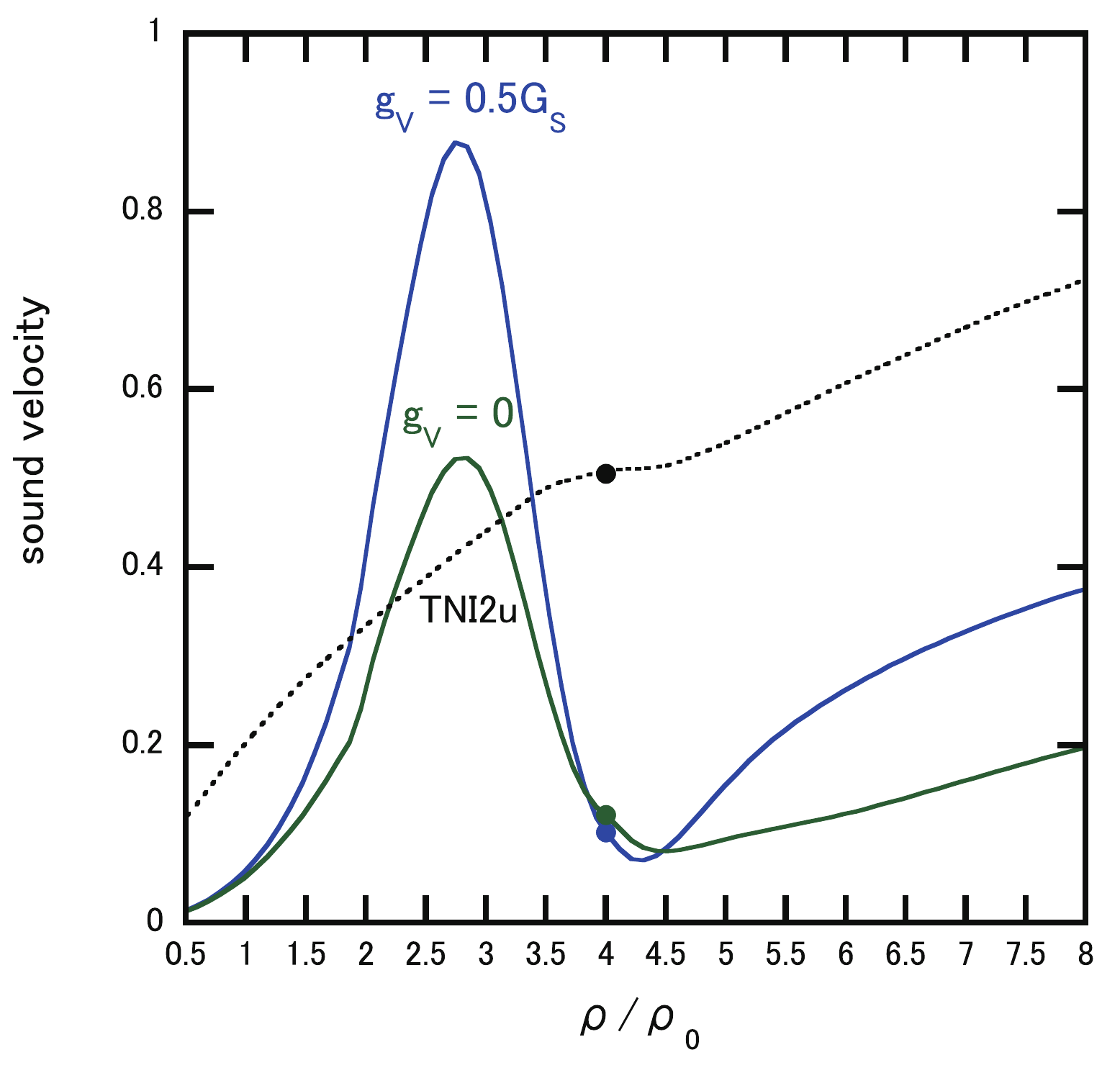}
 \caption{\footnotesize{
Sound velocity $v_{_S}$ as a function of baryon density $\rho$.
Solid lines: $v_{_S}-\rho$ obtained from the
interpolated EOS between the H-EOS with TNI2u and the Q-EOS with $g_{_V}/G_{_S}=0, 0.5$.
The crossover window is  $(2-4)\rho_0$. 
 Dotted line : that for  pure H-EOS with TNI2u. 
 The filled circles denote the points beyond which strangeness starts to appear.
}}
      \label{sound velocity-energy}
\end{tabular}
\end{figure}

\begin{figure}[!h]
  \begin{center}
      \centerline{\includegraphics[scale=0.6]
                                     {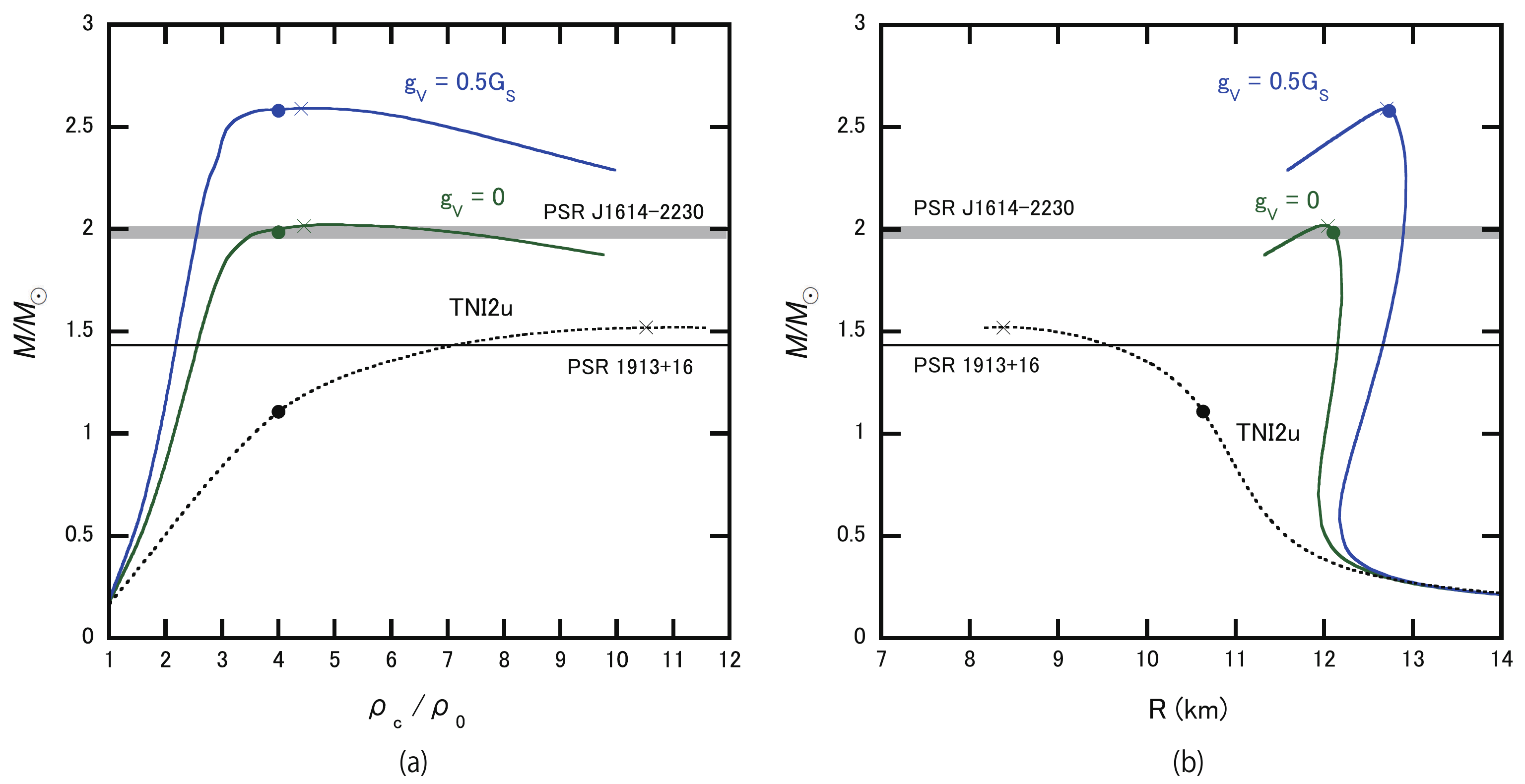}}
\end{center}
      \caption{\footnotesize{
(a) $M-\rho_c$ relationships with the interpolated EOSs.
We adopt  the  HK-parameter set for the Q-EOS with various  $g_{_V}/G_{_S}=0, 0.5$.
 The crossover window  are fixed to be $(\bar{\rho},\Gamma)=(3\rho_0,\rho_0)$.
The cross symbols denote the points of $M_{\rm max}$, while 
the filled circles denote the points beyond which the strangeness appears.
The gray band denotes $M=(1.97\pm0.04)M_{\odot}$ for PSR J1614-2230.
 The solid black line denotes $M=1.44M_{\odot}$ for PSR 1913+16.  
(b) $M$-$R$ relationships with the interpolated EOSs.  
 }}
      \label{m-rho-tni2u-energy}
 \end{figure} 
  
  Shown in Fig.\ref{interpolate-e-tni2u-energy} is the energy density
  interpolated between  TNI2u for H-EOS and NJL with $g_{_V}=0.5G_{_S}$ for Q-EOS.
  The crossover window is chosen to be  $(\bar{\rho}, \Gamma) = (3\rho_0, \rho_0)$
   and is shown by the shaded area on the horizontal axis.
  The pressure obtained from the interpolated energy density using 
  the thermodynamic relation is shown in   Fig.\ref{interpolate-p-tni2u-energy}.
  Due to the extra positive term $\Delta P$ in Eq.(\ref{eq:Delta-P}),
  the full pressure is larger than  $P_{H}$ and $P_{Q}$ in the 
   crossover region with the $\varepsilon$-interpolation
  procedure. Although $\Delta P$ is necessary for the thermodynamic consistency,
  its physical interpretation is not clear at the moment and is left for the future studies.  
 In Fig.\ref{interpolate-pe-tni2u-energy} and
  Fig.\ref{sound velocity-energy},
  we show $P$ as a function of $\varepsilon$ and the sound velocity $v_{_S}$ as a function of $\rho$,
   respectively.
  Because of the effect of $\Delta P$, the EOS becomes  stiff 
   and $v_{_S}$ is enhanced particularly in the crossover region. 
  Thus the maximum mass of the neutron star would become large even for
   moderate value of $g_{_V}$. 

 In Fig.\ref{m-rho-tni2u-energy}(a),
 we plot $M-\rho_c$ relationship between TNI2u for H-EOS and NJL Q-EOS for $g_{_V}/G_{_S}=0, 0.5$ with HK parameter set.
 We choose the crossover window as $(\bar{\rho},\Gamma)=(3\rho_0,\rho_0)$.
  For comparison, the $M-\rho_c$ relationship only with TNI2u   is
 plotted by the dashed line.  
Fig.\ref{m-rho-tni2u}(b) shows the corresponding $M-R$ relations.
As anticipated from Fig. \ref{sound velocity-energy}, the maximum mass is larger than the case of the 
$P$-interpolation for given $g_{_V}$.
 
 In Table \ref{table 7}, we show  $M_{\rm max}$ and $\rho_c$
 for different  H-EOS, vector type interaction $g_{_V}$ and choice of the crossover window parameterized by 
  $\bar{\rho}$ and $\Gamma$. 
The $\varepsilon$-interpolation makes EOS stiff more drastically than the $P$-interpolation. 
Even for $(g_{_V}, \bar{\rho})=(0,3\rho_0)$ and
$(g_{_V}, \bar{\rho})=(0.5, 5\rho_0)$, the maximum mass $M_{{\rm max}}$ can exceed
  $1.97M_{\odot}$. 
   
\begin{table}[!h]
\caption{\footnotesize{
$M_{\rm max}/M_{\odot}$ ($\rho_c/\rho_0$) for 
 different choice of H-EOS, stiffness of Q-EOS and crossover window.
 }} 
\label{table 7}
\begin{center}
  \begin{tabular}[c]{c|c|c|c|c}  \hline \hline
 & \multicolumn{2}{c|}{$g_{_V}=0$} &\multicolumn{2}{|c}{$g_{_V}=0.5G_{_S}$}\\
\cline{2-3}  \cline{3-4} \cline{4-5}
   H-EOS& $(\bar{\rho},\Gamma)=(3\rho_0,\rho_0)$ & $(5\rho_0,2\rho_0)$ & $(3\rho_0,\rho_0)$ & $(5\rho_0,2\rho_0)$ \\   \hline 
 TNI2u   & 2.02 (4.5) & 1.86 (8.7) & 2.59 (4.4) & 2.25 (6.1)   \\ 
 TNI2    & 2.02 (5.8) & 1.84 (9.1)  & 2.59 (4.3) & 2.23 (6.8)   \\ 
 TNI3u   & 1.99 (4.8) & 1.89 (8.5) & 2.57 (4.7) & 2.26 (6.0)   \\ 
 TNI3    & 1.97 (5.8) & 1.80 (6.3)  & 2.55 (4.5) & 2.21 (7.3) \\
Paris+TBF & 1.92 (4.8) & 1.75 (6.5) & 2.52 (4.7) & 2.17 (6.5)  \\
 AV18+TBF & 1.94 (4.7) & 1.75 (7.2) & 2.53 (4.7)& 2.19  (6.1)    \\  
 SCL3$\Lambda \Sigma$& 1.85 (4.8)& 1.73 (7.7)   & 2.46 (4.7)& 2.15 (6.8)   \\ 
\hline \hline
\end{tabular} 
\end{center}
\end{table}

\section{Summary and concluding remarks}

Recent observation of a two-solar mass NS presents a challenging problem how to reconcile  stiff EOS suggested from the observational side with  soft EOS due to hyperon-mixing  from the
 theoretical side.  
In this paper we have studied this problem on the basis of the percolation picture from the hadronic matter with hyperons to the quark matter with strange quarks.  We have constructed an EOS  by the interpolation between the H-EOS at lower densities and the Q-EOS at higher densities, and 
 found that the hybrid stars  could have $M_{{\rm max}} \sim 2M_{\odot}$, compatible with the observation.  This conclusion is in  contrast to the conventional EOS for hybrid stars derived through the Gibbs construction in which the resultant EOS becomes always softer than hadronic EOS and thereby leads to smaller $M_{{\rm max}}$.

  Our qualitative conclusion is insensitive to the choice of different types of 
  H-EOS and  different types of vector interaction  in Q-EOS
   as far as (i) the crossover 
  between the hadronic matter and the quark matter proceeds in a relatively low density region,
  (${\rho}=(2-4)\rho_0$), and  (ii) the quark matter is strongly interacting and stiff
    ($g_{_V}/G_{_S} \sim 1$).  These conditions applied to the $P$-interpolation procedure
  can be relaxed further if $\varepsilon$-interpolation is adopted. 
     We found that the 
  the sound velocity $v_{_S}$, which increases rapidly in the crossover window for 
  $g_{_V}/G_{_S} \geq 1$, 
     can nicely characterize the stiffening of the interpolated EOS and associated
     enhancement of $M_{{\rm max}}$. 
%   By increasing $g_{_V}$ further, we can have hybrid stars close to the causality limit,
%    $M_{{\rm max}} \sim 2.4 M_{\odot}$.
        
 The idea of rapid stiffening of the EOS starting from  $2\rho_0$ opens
 a possibility that the experimental nuclear incompressibility $\kappa=(240\pm20)$MeV
  at $\rho \sim \rho_0$ is compatible with the existence of massive neutron stars.
  Also, the idea  may well
 be checked by independent laboratory experiments with medium-energy heavy-ion collisions. 
 
 Although the $M$-$R$ relationship and $M_{\rm max}$ are insensitive to the 
  existence of the universal three-body repulsion,
 the onset density of strangeness is rather sensitive to such repulsion.
 If we have three-body repulsion acting universally among baryons,
most of the hybrid stars with $M \leq (1.9-2.0)M_{\odot}$ are free from the extremely
efficient hyperon direct-Urca cooling process and can avoid
  contradiction to observations.

 Finally, we remark that the crossover region may contain richer non-perturbative phases such as color superconductivity, inhomogeneous structures and so on \cite{fukushima-hatsuda}.
   How these structures as well as
  the  associated cooling processes affect the results of the present paper 
  would be an interesting future problem to be examined.

\section*{Acknowledgment}

We thank Wolfram Weise, Gordon Baym and David Blaschke, Mark Alford for discussions.
T.T. thanks Ryozo Tamagaki, Toshitaka Tatsumi and Shigeru Nishizaki for discussions and interests in this work. We also thank K. Tsubakihara and A. Ohnishi for providing us with the numerical
 data of the SCL3 EOS.
This  research was  supported in  part by  MEXT Grant-in-Aid  for Scientific
Research  on  Innovative  Areas(No.2004:20105003) by JSPS
Grant-in-Aid  for Scientific   Research (B)  No.22340052,
and by RIKEN 2012 Strategic Programs for R \& D.\\

\clearpage

\appendix
\def\thesection{Appendix\Alph{section}}

\section{EOS tables}

In this Appendix, we show the concrete values of pressure $P$ and energy density $\varepsilon$ as a function of a baryon density $x \equiv \rho/\rho_0$ for H-EOSs, Q-EOSs and interpolated EOSs.\\
In Table A1, we show H-EOSs with hyperons; TNI2, TNI2u, TNI3, TNI3u \cite{Nishizaki_2001,Nishizaki_2002}, AV18+TBF \cite{Baldo} and SCL3$\Lambda \Sigma$ \cite{Tsubakihara}. \\
Shown in Table A2, NJL Q-EOSs with HK parameter set for various vector interactions $g_{_V}/G_{_S}=0,1.0,1.5$ are listed in each row \cite{Hatsuda-Kunihiro}.\\
In Table A3, EOSs obtained by the interpolation of pressure between TNI2u H-EOS and NJL Q EOS with HK parameter set with 
$g_{_V}/G_{_S}=0, 1.0, 1.5$ for $(\bar{\rho}, \Gamma) = (3\rho_0, \rho_0)$ are listed.\\
In Table A4, EOSs obtained by the interpolation of energy density between TNI2u H-EOS and NJL Q EOS with HK parameter set with 
$g_{_V}/G_{_S}=0, 1.0, 1.5$ for $(\bar{\rho}, \Gamma) = (3\rho_0, \rho_0)$ are listed.

\begin{table}[!h] 
\caption{\footnotesize{Pressure $P$(MeV/fm$^3$) and energy density $\varepsilon$(MeV/fm$^3$) as a function of a baryon density $x \equiv \rho/\rho_0$ for various hadronic EOSs with hyperons\cite{Nishizaki_2001,Nishizaki_2002,Baldo,Tsubakihara}.} }
\begin{center}
  \begin{tabular}[c]{c c|c|c|c|c|c|c|c}  \hline \hline
\multicolumn{1}{c|}{} & \multicolumn{2}{|c|}{TNI2} & \multicolumn{2}{|c|}{TNI3} & \multicolumn{2}{|c|}{TNI2u} & \multicolumn{2}{|c}{TNI3u} \\ 
 \cline{2-3}  \cline{3-4}\cline{4-5}\cline{5-6}\cline{6-7}\cline{7-8}\cline{8-9}
 \multicolumn{1}{c|}{$x$}   
 & P
 & $\varepsilon$
 & P
 & $\varepsilon$
 & P
 & $\varepsilon$
 & P
 & $\varepsilon$
  \\ \hline 
\multicolumn{1}{c|}{1.0} & 2  & 162  & 3   &162  &2   &162 &3    &162   \\
\multicolumn{1}{c|}{ 1.5} & 7  & 245  & 9   &246  &7   &245 &9    &246   \\
\multicolumn{1}{c|}{ 2.0} & 15 & 330  & 20  &332  &15  &330 &20   &332   \\
\multicolumn{1}{c|}{ 2.5} & 26 & 418  & 32  &422  &26  &418 &38   &422   \\
\multicolumn{1}{c|}{ 3.0} & 40 & 508  & 40  &513  &42  &508 &61   &516   \\
\multicolumn{1}{c|}{ 3.5} & 48 & 600  & 48  &606  &62  &601 &92   &615   \\
\multicolumn{1}{c|}{ 4.0} & 57 & 693  & 58  &700  &87  &697 &131  &718   \\
\multicolumn{1}{c|}{ 4.5} & 67 & 787  & 69  &795  &112 &797 &174  &827   \\
\multicolumn{1}{c|}{ 5.0} & 80 & 882  & 82  &892  &140 &899 &224  &941   \\
\multicolumn{1}{c|}{ 5.5} & 93 & 979  & 96  &990  &173 &1005&283  &1060  \\
\multicolumn{1}{c|}{ 6.0} & 109& 1078 & 113 &1090 &211 &1113&353  &1185  \\
\multicolumn{1}{c|}{ 6.5} & 127& 1177 & 131 &1191 &255 &1226&433  &1316  \\
\multicolumn{1}{c|}{ 7.0} & 147& 1278 & 152 &1293 &304 &1341&525  &1454  \\
\multicolumn{1}{c|}{ 7.5} & 169& 1381 & 175 &1397 &360 &1461&629  &1599  \\
\multicolumn{1}{c|}{ 8.0} & 193& 1485 & 200 &1503 &422 &1584&746  &1751  \\
\multicolumn{1}{c|}{ 8.5} & 219& 1590 & 227 &1610 &492 &1712&876  &1911  \\
\multicolumn{1}{c|}{ 9.0} & 248& 1698 & 256 &1719 &568 &1843&1020 &2079  \\
\multicolumn{1}{c|}{ 9.5} & 278& 1807 & 287 &1829 &651 &1979&1178 &2256  \\
\multicolumn{1}{c|}{ 10.0}& 311& 1917 & 321 &1942 &743 &2120&1352 &2441  \\
\hline \hline
\\
\end{tabular} 

  \begin{tabular}[c]{c|c|c|c|c|c|c|c}  \hline \hline
  & \multicolumn{2}{|c|}{Paris+TBF} & \multicolumn{2}{|c|}{AV18+TBF} & \multicolumn{3}{|c}{SCL3$\Lambda \Sigma$}\\ 
 \cline{2-3}  \cline{3-4}\cline{4-5}\cline{5-6}\cline{6-7}\cline{7-8}
   $x$  
 & P
 & $\varepsilon$
 & P
 & $\varepsilon$
 & $x$  
 & P
 & $\varepsilon$  \\ \hline 
  0.471& 0.370& 71.2& 0.432  & 71.2& 1.02&5.85&167   \\ 
  0.941& 2.41 & 148 & 2.59   & 148 & 1.51&14.6&251   \\ 
   1.18& 4.63 & 187 & 4.94   & 187 & 2.04&28.9&346   \\ 
   1.76& 15.5 & 287 & 15.5   & 288 & 2.51&41.5&434   \\
   2.35& 31.2 & 404 & 29.0   & 399 & 3.02&53.5&531   \\ 
   2.94& 45.7 & 503 & 44.2   & 505 & 3.54&67.8&635   \\ 
   3.53& 62.3 & 617 & 59.9   & 623 & 4.07&84.0&740   \\ 
   4.12& 79.0 & 735 & 75.3   & 729 & 4.57&101 &841   \\
   4.71& 99.9 & 853 & 94.4   & 858 & 5.01&118 &933   \\ 
   5.29& 117  & 965 & 112    & 970 & 5.62&144 &1063  \\ 
   5.88& 145  & 1128& 139    & 1122& 6.02&162 &1150  \\ 
   6.47& 168  & 1240& 159    & 1223& 6.60&191 &1278  \\
   7.06& 188  & 1369& 181    & 1341& 7.07&216 &1384  \\ 
   7.65& 213  & 1470& 205    & 1459& 7.58&244 &1499  \\ 
   8.24& 242  & 1599& 239    & 1616& 8.12&276 &1625  \\ 
   8.82& 279  & 1745& 270    & 1745& 8.50&300 &1715  \\
   9.41& 307  & 1874& 302    & 1879& 9.11&339 &1860  \\ 
   10.0& 347  & 2031& 328    & 1986& 9.54&368 &1965  \\ 
\hline \hline
\end{tabular} 

\end{center}
\end{table}

\begin{table}[!h] 
\caption{\footnotesize{Pressure $P$(MeV/fm$^3$) and energy density $\varepsilon$(MeV/fm$^3$) as a function of a baryon density $x \equiv \rho/\rho_0$ for NJL Q-EOSs with HK parameter set for $g_{_V}=G_{_S}$ \cite{Hatsuda-Kunihiro}. }}
\begin{center}
  \begin{tabular}[c]{c|c|c|c|c|c|c}  \hline \hline
 & \multicolumn{2}{|c|}{$g_{_V}/G_{_S}=0$} & \multicolumn{2}{|c|}{$g_{_V}/G_{_S}=1$} & \multicolumn{2}{|c}{$g_{_V}/G_{_S}=1.5$} \\ 
 \cline{2-3}  \cline{3-4}\cline{4-5}\cline{5-6}\cline{6-7}
   $x$  
 & P
 & $\varepsilon$
 & P
 & $\varepsilon$
 & P
 & $\varepsilon$  \\ \hline 
 1.0 & -0.7633 & 179.2  & 8.390 &188.3 &12.97  &192.9     \\ 
 1.5 & -1.397  & 268.1  & 19.20 &288.7 &29.49  &299.0     \\ 
 2.0 & 6.721   & 357.8  & 43.33 &394.5 &61.64  &412.8     \\
 2.5 & 28.71   & 451.3  & 85.91 &508.5 &114.5  &537.1     \\ 
 3.0 & 58.56   & 550.0  & 140.9 &632.4 &182.1  &673.6     \\ 
 3.5 & 91.96   & 654.0  & 204.1 &766.2 &260.1  &822.2     \\ 
 4.0 & 127.5   & 762.8  & 274.0 &909.3 &347.2  &982.5     \\
 4.5 & 158.2   & 876.1  & 343.6 &1061  &436.2  &1154     \\ 
 5.0 & 182.8   & 992.3  & 411.6 &1221  &526.0  &1335     \\ 
 5.5 & 203.0   & 1111   & 479.9 &1388  &618.3  &1526     \\ 
 6.0 & 221.0   & 1231   & 550.5 &1561  &715.2  &1725     \\
 6.5 & 238.9   & 1353   & 625.6 &1740  &819.0  &1933     \\ 
 7.0 & 258.3   & 1476   & 706.8 &1925  &931.0  &2149     \\ 
 7.5 & 279.8   & 1601   & 794.7 &2115  &1052   &2373     \\ 
 8.0 & 303.7   & 1727   & 889.5 &2313  &1182   &2605     \\
 8.5 & 330.0   & 1854   & 991.3 &2516  &1322   &2846     \\ 
 9.0 & 358.4   & 1984   & 1100  &2725  &1471   &3096     \\ 
 9.5 & 389.9   & 2114   & 1215  &2941  &1628   &3354     \\ 
 10&   421.3   & 2247   & 1337  &3163  &1794   &3620     \\
\hline \hline
\end{tabular} 
\end{center}
\end{table}

\begin{table}[!h] 
\caption{\footnotesize{Pressure $P$(MeV/fm$^3$) and energy density $\varepsilon$(MeV/fm$^3$) as a function of a baryon density $x \equiv \rho/\rho_0$ for EOSs obtained by the interpolation of pressure between TNI2u H-EOS and NJL Q EOS with HK parameter set with $g_{_V}/G_{_S}=0, 1.0, 1.5$ for $(\bar{\rho}, \Gamma) = (3\rho_0, \rho_0)$ case. For the practical use, in a hadronic phase, we fit pressure as a function $a\rho^b$ by using the least-squares method.} }
\begin{center}
  \begin{tabular}[c]{c|c|c|c|c|c|c|c|c|c|c|c|c}  \hline \hline
 & \multicolumn{6}{|c|}{TNI2} & \multicolumn{6}{|c}{TNI2u} \\ 
 \cline{2-3}  \cline{3-4}\cline{4-5}\cline{5-6}\cline{6-7}\cline{6-7}
\cline{7-8}\cline{8-9}\cline{9-10}\cline{10-11}\cline{11-12}\cline{12-13}
 & \multicolumn{2}{|c|}{$g_{_V}/G_s=0$} & \multicolumn{2}{|c|}{$g_{_V}/G_s=1$}
& \multicolumn{2}{|c|}{$g_{_V}/G_s=1.5$} & \multicolumn{2}{|c|}{$g_{_V}/G_s=0$}
& \multicolumn{2}{|c|}{$g_{_V}/G_s=1$} & \multicolumn{2}{|c}{$g_{_V}/G_s=1.5$} \\ 
 \cline{2-3}  \cline{3-4}\cline{4-5}\cline{5-6}\cline{6-7}
\cline{7-8}\cline{8-9}\cline{9-10}\cline{10-11}\cline{11-12}\cline{12-13}
   $x$  
 & P
 & $\varepsilon$
 & P
 & $\varepsilon$
 & P
 & $\varepsilon$
 & P
 & $\varepsilon$
 & P
 & $\varepsilon$  
 & P
 & $\varepsilon$  \\ \hline 
 1.0 & 2.480 & 168.9 & 2.645 &179.5 &2.727 &184.8 &2.480 &168.8  & 2.645 &179.4 & 2.727 & 184.8   \\ 
 1.5 & 6.737 & 255.4 & 7.714 &271.5 &8.202 &279.6 &6.737 &255.3  & 7.714 &271.4 & 8.202 & 279.5   \\ 
 2.0 & 13.90 & 343.7 & 18.26 &366.0 &20.45 &377.1 &13.90 &343.6  & 18.26 &365.9 & 20.45 & 377.0   \\ 
 2.5 & 26.93 & 434.4 & 42.31 &464.4 &50.00 &479.4 &26.93 &434.3  & 42.31 &464.3 & 50.00 & 479.3   \\
 3.0 & 49.03 & 528.1 & 90.22 &569.3 &110.8 &589.8 &50.19 &528.5  & 91.38 &569.7 & 112.0 & 590.3   \\ 
 3.5 & 80.26 & 629.2 & 162.2 &687.2 &203.2 &716.2 &83.90 &627.5  & 165.8 &685.5 & 206.8 & 714.5   \\ 
 4.0 & 119.2 & 733.0 & 248.2 &814.2 &312.7 &854.8 &122.7 &731.6  & 251.7 &812.8 & 316.2 & 853.4   \\ 
 4.5 & 153.9 & 841.7 & 330.4 &952.0 &418.7 &1007 &156.0 &841.7  & 332.6 &952.0 & 420.9 & 1007    \\
 5.0 & 181.0 & 953.8 & 405.7 &1099  &518.0 &1171 &182.1 &954.0  & 406.8 &1099  & 519.1 & 1171    \\ 
 5.5 & 202.3 & 1068  & 477.3 &1252  &614.8 &1344 &202.8 &1069   & 477.8 &1253  & 615.4 & 1345    \\ 
 6.0 & 220.7 & 1185  & 549.4 &1413  &713.7 &1527 &220.9 &1185   & 549.6 &1413  & 714.0 & 1527    \\ 
 6.5 & 238.8 & 1303  & 625.1 &1579  &818.3 &1718 &238.9 &1303   & 625.3 &1580  & 818.4 & 1718    \\
 7.0 & 258.2 & 1422  & 706.6 &1752  &930.7 &1917 &258.3 &1422   & 706.6 &1752  & 930.8 & 1918    \\ 
 7.5 & 279.8 & 1543  & 794.6 &1931  &1052  &2125 &279.8 &1543   & 794.6 &1931  & 1052  & 2125    \\  
 8.0 & 303.7 & 1665  & 889.5 &2115  &1182  &2341 &303.7 &1665   & 889.5 &2116  & 1182  & 2341    \\ 
 8.5 & 330.0 & 1789  & 991.3 &2306  &1322  &2565 &330.0 &1789   & 991.3 &2307  & 1322  & 2566    \\
 9.0 & 358.4 & 1914  & 1100 &2503   &1471  &2798 &358.4 &1915   & 1100  &2504  & 1471  & 2799    \\ 
 9.5 & 389.0 & 2041  & 1215 &2707   &1628  &3039 &388.9 &2042   & 1215  &2707  & 1628  & 3040    \\ 
 10.0& 421.3 & 2170  & 1337 &2916   &1794  &3289 &421.3 &2170   & 1337  &2917  & 1794  & 3290    \\ 
\hline \hline
\end{tabular} 
\end{center}
\end{table}

\begin{table}[!h] 
\caption{\footnotesize{Pressure $P$(MeV/fm$^3$) and energy density $\varepsilon$(MeV/fm$^3$) as a function of a baryon density $x \equiv \rho/\rho_0$ for EOSs obtained by the interpolation of energy density between TNI2u H-EOS and NJL Q EOS with HK parameter set with $g_{_V}/G_{_S}=0, 0.5$ for $(\bar{\rho}, \Gamma) = (3\rho_0, \rho_0)$ case.} }
\begin{center}
  \begin{tabular}[c]{c|c|c|c|c|c|c|c|c}  \hline \hline
 & \multicolumn{4}{|c|}{TNI2} & \multicolumn{4}{|c}{TNI2u} \\ 
 \cline{2-3}  \cline{3-4}\cline{4-5}\cline{5-6}\cline{6-7}\cline{6-7}
\cline{7-8}\cline{8-9}
 & \multicolumn{2}{|c|}{$g_{_V}/G_s=0$} & \multicolumn{2}{|c|}{$g_{_V}/G_s=0.5$}
 & \multicolumn{2}{|c|}{$g_{_V}/G_s=0$} & \multicolumn{2}{|c|}{$g_{_V}/G_s=0.5$} \\ 
 \cline{2-3}  \cline{3-4}\cline{4-5}\cline{5-6}\cline{6-7}
\cline{7-8}\cline{8-9}
   $x$  
 & P
 & $\varepsilon$
 & P
 & $\varepsilon$
 & P
 & $\varepsilon$
 & P
 & $\varepsilon$ \\ \hline 
 1.0 & 2.844 & 162.3 & 3.088 &162.4 &2.844 &162.3 &3.088 &162.4 \\% & 2.645 &179.4 & 2.727 & 184.8   \\ 
 1.5 & 9.940 & 245.1 & 11.82 &245.6 &9.941 &245.1 &11.83 &245.6 \\% & 7.714 &271.4 & 8.202 & 279.5   \\ 
 2.0 & 25.71 & 333.3 & 35.59 &335.5 &25.72 &333.3 &35.59 &335.5 \\% & 18.26 &365.9 & 20.45 & 377.0   \\ 
 2.5 & 60.74 & 426.2 & 96.55 &433.9 &60.74 &426.2 &96.55 &433.9 \\% & 42.31 &464.3 & 50.00 & 479.3   \\
 3.0 & 113.6 & 528.5 & 196.0 &549.1 &113.2 &529.0 &195.6 &549.6 \\% & 91.38 &569.7 & 112.0 & 590.3   \\ 
 3.5 & 155.8 & 639.2 & 274.0 &680.2 &156.9 &639.8 &275.0 &680.8 \\% & 165.8 &685.5 & 206.8 & 714.5   \\ 
 4.0 & 178.6 & 754.4 & 304.6 &818.9 &177.2 &755.1 &303.2 &819.6 \\% & 251.7 &812.8 & 316.2 & 853.4   \\ 
 4.5 & 190.5 & 871.8 & 316.5 &960.1 &188.2 &872.3 &314.1 &960.6  \\%& 332.6 &952.0 & 420.9 & 1007    \\
 5.0 & 200.5 & 990.4 & 333.0 &1103  &198.4 &990.7 &330.9 &1103  \\%& 406.8 &1099  & 519.1 & 1171    \\ 
 5.5 & 211.9 & 1109  & 359.6 &1248  &210.5 &1110 &358.1&1248 \\% & 477.8 &1253  & 615.4 & 1345    \\ 
 6.0 & 225.3 & 1231  & 394.5 &1395  &224.4 &1231 &393.6 &1395  \\% & 549.6 &1413  & 714.0 & 1527    \\ 
 6.5 & 240.9 & 1353  & 436.4 &1546  &240.4 &1353 &435.9 &1546  \\% & 625.3 &1580  & 818.4 & 1718    \\
 7.0 & 259.2 & 1476  & 484.4 &1700  &258.9 &1476 &484.1 &1700  \\% & 706.6 &1752  & 930.8 & 1918    \\ 
 7.5 & 280.2 & 1601  & 538.1 &1858  &280.1 &1601 &538.0 &1858 \\%  & 794.6 &1931  & 1052  & 2125    \\  
 8.0 & 303.9 & 1727  & 597.0 &2020  &303.8  &1727 &596.9 &2020  \\% & 889.5 &2116  & 1182  & 2341    \\ 
 8.5 & 330.1 & 1854  & 660.8 &2185  &330.0  &1854 &660.8 &2185 \\%  & 991.3 &2307  & 1322  & 2566    \\
 9.0 & 358.5 & 1984  & 729.2&2354  &358.5  &1984&729.2 &2354\\%  & 1100  &2504  & 1471  & 2799    \\ 
 9.5 & 389.0 & 2115  & 802.0 &2528   &389.0  &2115 &802.0 &2528 \\%  & 1215  &2707  & 1628  & 3040    \\ 
 10.0& 421.3 & 2247  & 879.0 &2705   &421.3  &2247 &878.9 &2705\\%   & 1337  &2917  & 1794  & 3290    \\ 
\hline \hline
\end{tabular} 
\end{center}
\end{table}

\vfill\pagebreak

\end{document}